%% file: main.tex
\documentclass{article}
\PassOptionsToPackage{authoryear,round}{natbib}
\usepackage{arxiv}
\usepackage{natbib}
\usepackage[utf8]{inputenc}
\usepackage[T1]{fontenc}
\usepackage{amsmath,amssymb,booktabs,array,longtable,graphicx}
\usepackage{microtype}
\usepackage{fvextra}
\usepackage{xurl}
\usepackage[hidelinks]{hyperref}
\usepackage{bookmark}
\usepackage{placeins}
\hypersetup{
  pdftitle={The Weight of Teamwork: Does Communication Structure Help LLM Scientists Search for Scientific Laws?},
  pdfauthor={Jianing Shi, Zaibin Zhang, Shijun Lei, Zhenfei Yin},
  pdfsubject={Communication structure and selection in multi-agent LLM scientific search},
  pdfkeywords={large language models, multi-agent systems, communication topology, symbolic regression, scientific discovery}
}
\newcommand{\tightlist}{\setlength{\itemsep}{0pt}\setlength{\parskip}{0pt}}
\newcommand{\tablefont}{\fontsize{8}{9.5}\selectfont}
\newcommand{\code}[1]{\texttt{\path{#1}}}

\title{The Weight of Teamwork:\\
Does Communication Structure Help LLM Scientists\\
Search for Scientific Laws?}
\author{
  Jianing Shi\textsuperscript{1} \quad
  Zaibin Zhang\textsuperscript{2} \quad
  Shijun Lei\textsuperscript{3} \quad
  Zhenfei Yin\textsuperscript{4,*}\\[0.45em]
  \textsuperscript{1}The University of Chicago (uchicago.edu) \quad
  \textsuperscript{2}Dalian University of Technology (dlut.edu.cn)\\
  \textsuperscript{3}Northwestern Polytechnical University (nwpu.edu.cn) \quad
  \textsuperscript{4}University of Oxford (ox.ac.uk)\\[0.2em]
  {\scriptsize\shortstack[c]{Emails: \textsuperscript{1}\texttt{jianingshi@uchicago.edu} \;
  \textsuperscript{2}\texttt{dlutzzb@gmail.com}\\
  \textsuperscript{3}\texttt{shijunlei@mail.nwpu.edu.cn} \;
  \textsuperscript{4}\texttt{jeremyyin@robots.ox.ac.uk} \quad
  \textsuperscript{*}Corresponding author: Zhenfei Yin}}
}
\date{}
\renewcommand{\headeright}{}
\renewcommand{\undertitle}{}
\renewcommand{\shorttitle}{The Weight of Teamwork}
\begin{document}
\maketitle
\begin{abstract}
\input{sections/abstract}
\end{abstract}
\keywords{large language models \and multi-agent systems \and communication topology \and symbolic regression \and scientific discovery}
\input{sections/main_body}
\FloatBarrier
\label{main-text-end}
\clearpage
\section*{Acknowledgements}

We thank the PhAI Lab for providing API access for part of the
experiments reported here. The remaining computational
experiments were run on local hardware.

We thank the anonymous reviewers at the NeurIPS 2026 FAST
workshop for comments that improved the presentation and the
scope of the claims.

\bibliographystyle{plainnat}
\bibliography{references}
\clearpage
\appendix
\renewcommand{\thetable}{\Alph{section}\arabic{table}}
\renewcommand{\thefigure}{\Alph{section}\arabic{figure}}
\renewcommand{\theHtable}{\Alph{section}.\arabic{table}}
\renewcommand{\theHfigure}{\Alph{section}.\arabic{figure}}
\input{sections/appendices}
\end{document}

%% file: sections/abstract.tex
Team performance is often reported as one delivered score, even though that score is produced in two stages: a system generates candidate answers and a selection rule returns one of them. We use symbolic regression as an observable scientific-search setting and compare fixed-budget LLM teams that differ in peer visibility and update timing, evaluating both their frozen candidate archives and the expressions selected from those archives. Distributing the call budget across more independent trajectories produced the largest measured gain, improving mean held-out fitting score \(Q\) by 0.70 at the widest tested allocation. Under a common archive-wide selector, the implemented Ring policy reduced mean \(Q\) by 0.33 relative to Independent, while Star and Complete remained unresolved. Holding the generated archives fixed, changing the delivery rule changed the measured architecture gaps, including the sign of the Complete point estimate. These results support reporting candidate generation and final selection as separate endpoints. They do not identify a communication mechanism or a universally preferred topology; rather, they show that an architecture ranking is not identified by communication structure alone when the rule that returns the final answer is left unspecified.

%% file: sections/main_body.tex
\section{Introduction}
Organizing an LLM team for scientific search involves choosing which ideas its members can see. Systems such as VirSci and Co-Scientist organize agents to propose, evaluate, and refine research ideas \citep{su2025many,gottweis2026coscientist}. A communication structure makes some hypotheses available and leaves others outside an agent's context. Sharing may support useful revision, but it may also expose several agents to the same unhelpful proposal. The research question is how this organization of information affects the hypotheses a team produces and the quality of its eventual answer.

The question of how communication structure affects group problem solving is not new. Classic five-person communication-network experiments found that fixed communication patterns changed task performance and member satisfaction \citep{bavelas1950communication,leavitt1951communication}. Those studies recorded written messages and encoded communication activity as well as the channels that were open. An LLM experiment can additionally preserve the exact system-mediated information made available to each scientist, every candidate expression, and every public rationale. This record does not reveal private cognition; it lets us separate what a team publicly generates from what its selection rule returns.

Knowing who can see a proposal does not establish what they do with it. A team may repeat a shared expression without improving its fit. It may also generate a better candidate that its final selection rule fails to return. The final score then leaves both the search process and selection partly hidden. We therefore examine communication through the generated archive as well as the returned expression. This separates the evidence for a policy's performance from the evidence needed to explain its mechanism.

We investigate this question through symbolic regression, where candidate expressions can be recorded and objectively evaluated. The task supplies a controlled instance of proposing and revisiting hypotheses from observations and prior discussion. The Qwen layer contains 100 problems, and the Terra layer contains 41 Transform problems. We specify peer visibility and update timing at a fixed invocation budget, then evaluate the resulting archives under shared and alternative selection rules.

In the exploratory Qwen comparisons, Ring reduced mean held-out \(Q\) by 0.33 relative to Independent (95\% CI {[}-0.49, -0.18{]}). Star's mean difference was +0.03 ({[}-0.08, +0.13{]}). Complete's was +0.10 ({[}-0.08, +0.34{]}). All three comparisons used the same archive-wide selector. The evidence supports a Ring disadvantage under common delivery, while the other intervals leave benefit and harm unresolved and do not establish equivalence.

As an allocation reference, ten independent trajectories improved mean held-out \(Q\) over one by 0.70 (95\% CI {[}0.41, 1.03{]}). The median gain was 0.11. There were 86 wins, 11 losses, and 3 ties. This comparison changes trajectory length and context resets together with trajectory count. It provides evidence for the tested allocation policy without isolating which component produced the gain.

Fixed-archive comparisons show delivery sensitivity in Qwen, while the corresponding Terra interactions remain unresolved because their estimates are imprecise. Complete's point-estimate gap changes sign under final-round selection. The archive-wide interval includes zero. Archive diagnostics additionally distinguish team-level concentration from repetition within an agent. These measurements help interpret team performance at the level of candidate generation and selection.

This study does not introduce a new solver, identify a communication mechanism, or establish a universally preferred topology. It treats symbolic regression as an observable scientific-search instance. The central comparison concerns candidate quality under implemented communication policies and a shared selector. The archive diagnostics and selection comparisons clarify how far those outcomes can be interpreted. The resulting evidence informs the evaluation of scientific-agent teams while leaving broader mechanism questions open.

Our contributions are threefold:
\begin{itemize}
\tightlist
\item We show that distributing a fixed invocation budget across more independent trajectories yields the largest measured improvement in this testbed.
\item We find that the implemented serial Ring policy underperforms Independent under a common archive-wide selector, while Star and Complete remain unresolved.
\item We demonstrate on frozen candidate archives that the measured gap between architectures changes with the rule used to return a final expression.
\end{itemize}

\section{Scientific-search testbed and communication policies}

We used LLM-SRBench \citep{shojaee2025srbench} to study repeated hypothesis proposals with objective candidate evaluation. A prompt supplies a target, variable descriptions, and five training rows, and requests a short scientific discussion followed by one expression. This gives an observable record of proposed hypotheses and peer discussion. The Qwen layer contains 50 LSR-Transform and 50 LSR-Synth problems; the Terra layer uses the common 41-problem Transform intersection. This instance covers expression proposal and revision under restricted information, rather than the full range of scientific inquiry.

Each condition received 50 scheduled calls per problem. The communication comparison fixes five agents and ten rounds; the Qwen independent-allocation comparison uses \((N,T)\in\{(1,50),(2,25),(5,10),(10,5)\}\). Qwen \citep{qwen2024report} used \texttt{qwen2.5:7b-instruct}, temperature 0.70, and a 512-token output cap. Terra used \texttt{gpt-5.6-terra}, temperature 1.00, \texttt{reasoning\_effort=none}, and a 2,048-token completion cap. These are within-layer comparisons at equal invocation counts, without matching token consumption or claiming controlled effect-size differences between models. Appendix~\ref{app:A} gives the complete prompt and settings.

\begin{figure}[htbp]
\centering
\includegraphics[width=0.98\linewidth]{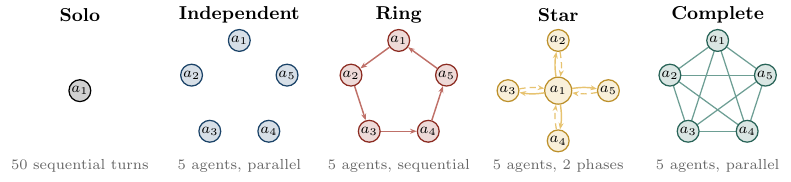}
\caption{Information available to each agent. Five panels show Solo, Independent, Ring, Star, and Complete. Arrows indicate whose responses a recipient can read. In Star, solid arrows run from hub to leaves and dashed arrows from leaves to hub. Ring uses the latest predecessor response in rotating serial order; Star updates leaves before the hub; Complete uses a previous-round snapshot. Solo and Independent have self-history without peer visibility. All panels represent 50 scheduled calls per problem. The policies vary visibility and update timing together, so graph connectivity alone does not specify the treatment.}
\label{fig:1}
\end{figure}

Figure~\ref{fig:1} specifies both visibility and timing. Independent receives no peer responses. Ring reads the latest predecessor response in rotating serial order. In Star, leaves read the latest hub response before the hub reads the current leaf responses. Complete reads a previous-round snapshot from all four peers. Each agent retains its own previous responses; peer text does not accumulate in that self-history. The prompt anonymizes colleagues and does not name the topology; the software supplies peer responses according to the policy. These implemented policies differ in update timing as well as connectivity.

We measured normalized mean squared error (NMSE) and its negative base-ten logarithm:

\[
\operatorname{NMSE}=\frac{n^{-1}\sum_i(\hat y_i-y_i)^2}{\operatorname{Var}_i(y_i)},
\qquad Q=-\log_{10}(\operatorname{NMSE}+10^{-10}).
\]

For nondegenerate targets, constant-mean prediction gives NMSE \(=1\) and \(Q\approx0\); positive \(Q\) indicates improvement over that reference up to the stabilizer. Differences in \(Q\) are logarithmic error differences. We use population variance over finite prediction--target pairs; a nonpositive variance triggers a denominator of \(\operatorname{mean}(y^2)\), then 1 if needed. The frozen stabilizer maps zero NMSE to \(Q=10\). Evaluation uses the official pregenerated held-out ID arrays, disjoint from training and never shown to the agents.

The \textbf{common-rule endpoint} evaluates each policy after the same archive-wide selector, \(d_1\), chooses the valid expression with lowest training NMSE. Validity requires a nonempty canonical skeleton and finite training NMSE; equal errors resolve to the lexicographically smallest message identifier. The \textbf{crossed-delivery endpoint} applies six prespecified selectors to the same archives; the focal emphasis on \(d_2\)--\(d_4\) is exploratory. The alternatives use final-round training fit, final-round or archive-wide plurality, the last message, or a complexity penalty. None uses held-out targets or the reference expression. No rule has a fallback; unavailable outputs are excluded through each contrast's common finite-score intersection. Appendix~\ref{app:B} defines the selectors and the per-problem delivery interaction.

Mean paired differences are primary. We report 95\% percentile problem-bootstrap intervals with 10,000 resamples and seed 20260904 \citep{efron1979bootstrap}, alongside medians, exact ties, signed-rank effect sizes, and Wilcoxon tests \citep{wilcoxon1945ranking}. Holm families \citep{holm1979multiple} contain the three pooled communication contrasts, five allocation contrasts, or 20 delivery interactions within a model layer. All pooled Qwen communication contrasts are exploratory because their family was documented after the results. Appendix~\ref{app:J} records that chronology, the endpoint amendment, and the historical Transform Wilcoxon convention with its full-vector sensitivity check.

\section{Communication and the returned result}

\subsection{Communication under a common selector}

In the exploratory Qwen comparisons, the implemented Ring policy reduced mean held-out \(Q\) by 0.33 relative to Independent (95\% CI {[}-0.49, -0.18{]}; Holm \(p=5.75\times10^{-7}\)). Star's mean difference was +0.03 ({[}-0.08, +0.13{]}; \(p=0.91\)). Complete's was +0.10 ({[}-0.08, +0.34{]}; \(p=0.36\)). Their median differences were -0.01, \(-5.64\times10^{-7}\), and \(+1.73\times10^{-5}\), respectively (Table~\ref{tab:main-contrasts}A). The latter intervals allow both benefit and harm, without establishing equivalence. With candidate selection held constant, these results support a disadvantage for the implemented Ring policy and leave the Star and Complete mean differences unresolved. Appendix~\ref{app:C} documents the split diagnostics used for these comparisons.

Ring's disadvantage also appeared within each Qwen task family: mean differences were -0.30 on Transform (95\% CI {[}-0.48, -0.14{]}) and -0.35 on Synth ({[}-0.63, -0.12{]}). Star's family-specific means were +0.11 and -0.06, with Holm \(p=0.29\) and 0.40. Complete's were +0.23 and -0.04, with \(p=0.21\) and 0.84. Neither Star nor Complete had a detected signed-rank difference from Independent within either family. Appendix~\ref{app:F} reports the six exported contrasts within each task family; Appendix~\ref{app:J} retains the additional Transform sensitivity. The change in their point ordering is descriptive and does not establish a task-family interaction.

Terra Transform provides a narrower comparison. Ring's mean difference was -1.18 (95\% CI {[}-2.04, -0.48{]}; Holm \(p=\ensuremath{1.63\times10^{-3}}\)). Star's mean was -0.57, with a zero-spanning interval ({[}-1.56, +0.40{]}) despite Holm \(p=0.03\) for the signed-rank test. Complete's mean was -0.36 ({[}-1.43, +0.73{]}; \(p=0.08\)). All three medians were zero, with 20--24 ties among 41 problems. The mean interval and signed-rank test describe different features of the paired differences; we did not test a model-by-policy interaction. Appendix~\ref{app:I} retains both summaries, so the Terra Star test result does not substitute for an estimate of its mean effect.

\begin{table}[t]
\caption{Communication comparisons and allocation reference. Mean held-out $Q$ differences and 95\% paired bootstrap intervals are primary. Medians, wins/losses/ties, signed-rank rank-biserial correlations ($r_{rb}$), and Holm-adjusted Wilcoxon values are complementary summaries. Panel A contains the three exploratory communication contrasts at $N=5$. Panel B varies $N$ without peer communication, retaining 50 calls per problem. Positive allocation differences favor the larger $N$. Appendix~\ref{app:F} gives the full stratified results.}
\label{tab:main-contrasts}
\centering\tablefont
\setlength{\tabcolsep}{2.0pt}
\par\smallskip\textbf{A. Qwen communication}\par\smallskip
\begin{tabular}{@{}>{\raggedright\arraybackslash}p{86.820pt}>{\raggedright\arraybackslash}p{113.244pt}>{\raggedright\arraybackslash}p{60.397pt}>{\raggedright\arraybackslash}p{37.748pt}>{\raggedright\arraybackslash}p{33.973pt}>{\raggedright\arraybackslash}p{45.298pt}@{}}\toprule
\textbf{Contrast} & \textbf{Mean $\Delta Q$ [95\% CI]} & \textbf{Median} & \textbf{W/L/T} & \textbf{$r_{rb}$} & \textbf{Holm $p$} \\
\midrule
Ring - Independent & -0.33 [-0.49, -0.18] & -0.01 & 21/75/4 & -0.61 & $5.75\times10^{-7}$ \\
Star - Independent & +0.03 [-0.08, +0.13] & $-5.64\times10^{-7}$ & 43/50/7 & -0.01 & 0.91 \\
Complete - Independent & +0.10 [-0.08, +0.34] & $+1.73\times10^{-5}$ & 53/42/5 & +0.16 & 0.36 \\
\bottomrule\end{tabular}

\centering\tablefont
\setlength{\tabcolsep}{2.0pt}
\par\smallskip\textbf{B. Qwen allocation}\par\smallskip
\begin{tabular}{@{}>{\raggedright\arraybackslash}p{86.820pt}>{\raggedright\arraybackslash}p{113.244pt}>{\raggedright\arraybackslash}p{60.397pt}>{\raggedright\arraybackslash}p{37.748pt}>{\raggedright\arraybackslash}p{33.973pt}>{\raggedright\arraybackslash}p{45.298pt}@{}}\toprule
\textbf{Allocation} & \textbf{Mean $\Delta Q$ [95\% CI]} & \textbf{Median} & \textbf{W/L/T} & \textbf{$r_{rb}$} & \textbf{Holm $p$} \\
\midrule
$1\to2$ & +0.23 [0.06, 0.43] & $6.30\times10^{-5}$ & 55/40/5 & 0.25 & 0.04 \\
$2\to5$ & +0.20 [0.07, 0.34] & \ensuremath{4.70\times10^{-3}} & 64/28/8 & 0.46 & $3.36\times10^{-4}$ \\
$5\to10$ & +0.26 [0.04, 0.58] & \ensuremath{5.20\times10^{-4}} & 58/37/5 & 0.33 & \ensuremath{9.46\times10^{-3}} \\
$1\to5$ & +0.43 [0.27, 0.63] & 0.05 & 78/18/4 & 0.74 & $1.08\times10^{-9}$ \\
$1\to10$ & +0.70 [0.41, 1.03] & 0.11 & 86/11/3 & 0.87 & $4.41\times10^{-13}$ \\
\bottomrule\end{tabular}
\end{table}

\subsection{A descriptive check on search behavior}

An exploratory archive diagnostic separates team-level concentration from repetition within an agent. A modal expression skeleton accounted for at least 80\% of a problem's outputs in 81 Solo cases, 72 Ring cases, 56 Star cases, 46 Complete cases, and no Independent cases. At the agent level, the corresponding proportion was 81\% for Solo and 81.80\% for Independent. Solo's best-so-far \(Q\) also improved by 0.01 between calls 10 and 50. Pooling different trajectories can reduce team-level concentration while individual trajectories remain repetitive, and improvement within a trajectory is still possible. These summaries constrain the interpretation of diversity but do not identify a peer-influence mechanism; Appendix~\ref{app:G} records the exploratory process analyses.

An exploratory archive summary records how many distinct canonical skeletons each condition generated per problem. The medians were 2 for Solo, 2 for Ring, 5 for Star, 7 for Complete, and 8.50 for Independent, with 3 at $N=2$ and 15.50 at $N=10$. Among the five-agent conditions, Ring had both the lowest median skeleton count and the fewest invalid candidate slots: 1 of 5,000, compared with 61 for Independent, 43 for Star, and 36 for Complete. Ring tied Solo on the reported median skeleton count; the invalid-slot definition also requires finite training NMSE and is broader than parse failure alone. These counts describe what each condition generated. They do not explain why the delivered scores differ: a post-hoc analysis relating within-problem changes in skeleton counts to changes in delivered $Q$ found no consistent association, and a stratified follow-up did not support the account that motivated it (Appendix~\ref{app:G}; extended function- and symbolic-space diagnostics appear in Appendix~\ref{app:H}).

\subsection{How selection changes the measured gap}

\begin{figure}[htbp]
\centering
\includegraphics[width=0.98\linewidth]{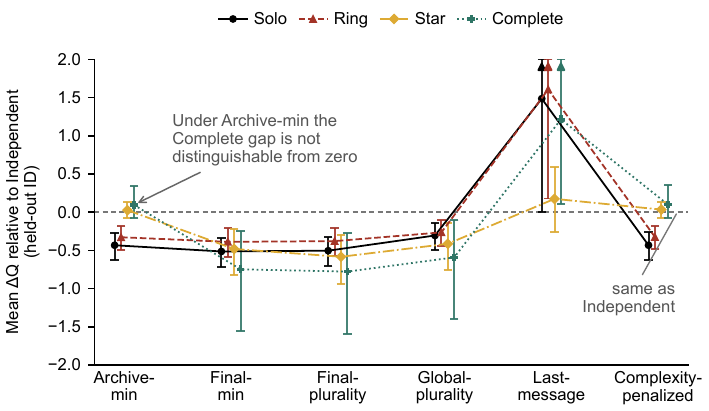}
\caption{Policy gaps across delivery rules. Points show Qwen mean held-out $Q$ differences from Independent for Solo, Ring, Star, and Complete under each rule; error bars are 95\% paired bootstrap intervals. Lines connect discrete rules and do not represent a continuous intervention. Both policies use the same rule on fixed archives. Valid pairs range from 96 to 100; Appendix~\ref{app:D} lists the cohorts. The vertical range is [-2, 2]. Outward arrows mark clipped $d_5$ upper limits: Solo 4.13, Ring 4.21, and Complete 3.12. The zero line is labeled ``same as Independent.'' Appendix~\ref{app:D} classifies sign changes using simple-effect uncertainty together with corrected interactions.}
\label{fig:2}
\end{figure}

With candidate archives fixed, 9/20 prespecified architecture-by-delivery contrasts passed Holm-adjusted signed-rank tests in Qwen. The Terra subset had 0/20, with estimates too imprecise to distinguish small interactions from no interaction. Complete's \(d_2\) interaction, for example, had mean -0.84 \(Q\) (95\% CI {[}-1.85, -0.26{]}; Holm \(p=1.05\times10^{-7}\)). The passing-test counts summarize signed ranks, not the number of mean intervals excluding zero. Since the generation record stays fixed, the comparison locates a dependence on selection in the measured architecture gap (Figure~\ref{fig:2}; Appendix~\ref{app:D}).

We examine \(d_2\), \(d_3\), and \(d_4\) as exploratory focal rules within that prespecified family. Under final-round minimum training error (\(d_2\)), Complete minus Independent had mean -0.75 \(Q\) (95\% CI {[}-1.56, -0.25{]}) and median \ensuremath{-6.10\times10^{-3}}. Under final-round plurality (\(d_3\)), its mean was -0.78 ({[}-1.59, -0.28{]}) and median \ensuremath{-9.58\times10^{-3}}. Archive-wide plurality (\(d_4\)) gave mean -0.59 ({[}-1.40, -0.10{]}) and median \ensuremath{-1.38\times10^{-3}}. The common rule (\(d_1\)) gave mean +0.10 ({[}-0.08, +0.34{]}) and median \(+1.73\times10^{-5}\). Because the \(d_1\) interval spans zero, the \(d_2\)/\(d_3\) cases remain point-estimate sign changes; Table~\ref{tab:delivery} and Appendix~\ref{app:D} separate those classifications from corrected interactions.

Complete's \(d_3\) interaction had Holm \(p=6.14\times10^{-8}\) with frozen skeletons and \(8.10\times10^{-8}\) with alternative functional canonicalization. Its \(d_4\) interaction failed Holm correction under both conventions (\(p=0.07\) and 0.16). The \(d_2\) result is unchanged because its selector does not group expressions, providing a grouping-independent delivery comparison (Appendix~\ref{app:D}).

\begin{table}[t]
\caption{Complete minus Independent under the focal delivery rules. All four rows use 100 paired Qwen problems. Means and intervals describe simple effects; the last column tests each rule’s change from $d_1$ using the prespecified family of 20 signed-rank interactions. The focal emphasis on $d_2$–$d_4$ is exploratory. The common-rule mean interval includes zero. Appendix~\ref{app:D} gives all architectures, rules, and sign-change classifications, including the supported Ring case under the weakest rule, the last-message stress test $d_5$.}
\label{tab:delivery}
\centering\tablefont
\setlength{\tabcolsep}{3.0pt}
\begin{tabular}{@{}>{\raggedright\arraybackslash}p{144.202pt}>{\raggedright\arraybackslash}p{121.434pt}>{\raggedright\arraybackslash}p{53.127pt}>{\raggedright\arraybackslash}p{60.717pt}@{}}\toprule
\textbf{Rule} & \textbf{Mean $\Delta Q$ [95\% CI]} & \textbf{Median} & \textbf{Interaction Holm $p$} \\
\midrule
$d_1$: archive minimum training error & +0.10 [-0.08, +0.34] & $+1.73\times10^{-5}$ & Reference \\
$d_2$: final-round minimum training error & -0.75 [-1.56, -0.25] & \ensuremath{-6.10\times10^{-3}} & $1.05\times10^{-7}$ \\
$d_3$: final-round plurality & -0.78 [-1.59, -0.28] & \ensuremath{-9.58\times10^{-3}} & $6.14\times10^{-8}$ \\
$d_4$: archive-wide plurality & -0.59 [-1.40, -0.10] & \ensuremath{-1.38\times10^{-3}} & 0.07 \\
\bottomrule\end{tabular}
\end{table}

\section{Independent exploration as a reference}

\subsection{Allocation and absolute performance}

At 50 calls per problem, allocating Qwen calls across ten independent trajectories instead of one, a best-of-\(N\) sampling idea related to self-consistency \citep{wang2023selfconsistency}, increased mean held-out \(Q\) by 0.70 (95\% CI {[}0.41, 1.03{]}). The median gain was 0.11, with 86 wins, 11 losses, and 3 ties. The \(N=1\to5\) reference gave mean gain 0.43 and median gain 0.05. These mean--median gaps indicate right-skewed gains. Because allocation also changes trajectory length, resets, and history dependence, the result supports the tested allocation policy as a whole (Appendix~\ref{app:E}).

\begin{figure}[htbp]
\centering
\includegraphics[width=0.62\linewidth]{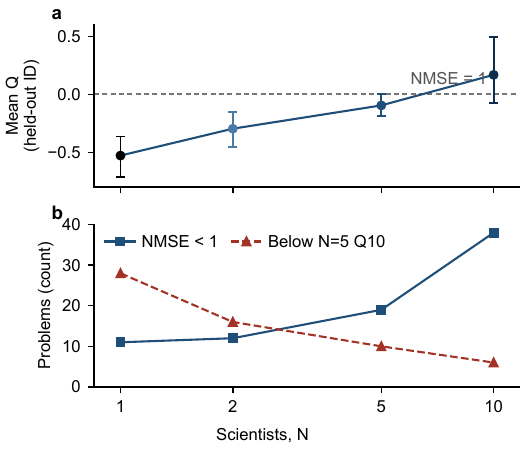}
\caption{Fixed-budget independent exploration. (A) Mean held-out $Q$ at $N=1,2,5,10$, with 95\% problem-bootstrap intervals and 50 calls per problem. The dashed line marks $Q=0$ (NMSE approximately 1). (B) Counts with NMSE $<1$ and an exploratory descriptive count below the strict lower-tail threshold $Q_{10}(N=5)=-0.76$. Selection uses archive-wide training fit. The positive paired gains do not imply that the absolute mean at $N=10$ exceeds zero with the same confidence.}
\label{fig:3}
\end{figure}

Each adjacent increase in \(N\) had a positive mean interval: the gains were 0.23, 0.20, and 0.26 \(Q\) units for \(1\to2\), \(2\to5\), and \(5\to10\). Their median gains were \(6.30\times10^{-5}\), \ensuremath{4.70\times10^{-3}}, and \ensuremath{5.20\times10^{-4}}. Exact ties numbered 5, 8, and 5. Small typical gains therefore coexist with changes on most problems; Figure~\ref{fig:3} and Appendix~\ref{app:E} place both features alongside the mean improvement.

Figure~\ref{fig:3} also reports absolute fitting performance. Mean \(Q\) at \(N=10\) was +0.17, with a 95\% interval spanning zero ({[}-0.08, +0.49{]}). At that allocation, 38/100 problems beat the constant-mean predictor, compared with 19/100 at \(N=5\). The \(N=10\) median was \ensuremath{-1.70\times10^{-3}}. The lower-tail counts in Figure~\ref{fig:3}B are exploratory descriptions, not an independent confirmation of the allocation contrast. These absolute measurements prevent the paired gain from being read as consistently successful fitting, even within this testbed (Appendix~\ref{app:E}).

The \(N=1\to5\) gain appeared in both Qwen families: mean 0.35 on Transform (95\% CI {[}0.16, 0.56{]}). On Synth, the mean was 0.52 ({[}0.26, 0.84{]}). Synth had median gain 0.04 and 44/6/0 wins/losses/ties. Terra Transform had mean gain 1.04 ({[}0.06, 2.07{]}), with median zero and 20 ties among 41 problems. Synth extends the result beyond transformed Feynman equations without testing memorization; the paired Terra Synth comparison was not completed. Appendices F and I preserve these scope and tie differences when interpreting the allocation reference.

\section{Discussion and scope}
\enlargethispage{2\baselineskip}

The communication comparison covers 100 Qwen problems and 41 Terra Transform problems. It concerns the organization of LLM hypothesis search in this instance, without establishing a general architecture for scientific discovery. Held-out fitting error measures candidate quality here; it does not measure the value of a scientific explanation. The paired Terra Synth comparison remains unavailable, and bootstrap intervals across problems do not measure repeated-run sampling variation. The historical Qwen runtime hash and SciPy version are unrecoverable; Appendix~\ref{app:J} states the provenance limits.

Prior work examines organization at different levels. \citet{wu2019teams} associate smaller human research teams with more disruptive contributions and larger teams with development of existing work. \citet{lazer2007network} use simulations to study how network structure changes collective search over time. \citet{kim2026societies} examine interacting perspectives within reasoning traces. These studies motivate the information-organization question, but their team-size associations, simulated dynamics, and internal reasoning processes are distinct from the external communication policies tested here.

Several lines of LLM-agent work optimize interaction itself. MacNet scales collaboration through a configurable communication network, AgentPrune removes low-value communication links, and multi-agent debate iterates critique and response to improve answer quality \citep{qian2025macnet,zhang2025agentprune,du2024debate,liang2024debate}. Scientific-agent systems automate proposal, evaluation, and refinement workflows \citep{su2025many,gottweis2026coscientist}. These studies optimize structure or workflow for task performance; we instead hold a small set of communication policies fixed and ask what a comparison among those policies measures once candidate generation and final selection are separated.

\citet{kim2026collaboration} compare single-agent and multi-agent systems across several benchmarks under matched reasoning-token budgets, and report that coordination gains diminish as single-agent capability increases. \citet{tran2026budgets} separately compare single- and multi-agent systems under equal thinking-token budgets. Kim et al.'s question is when coordination helps. Ours is how the measured gap between architectures depends on visibility and on the rule that returns a final answer. We do not apply their capability threshold to this testbed, because their leave-one-domain-out validation does not support transfer to a new task family.

Within scientific-agent systems, VirSci also compares sequential and random speaking modes \citep[Appendix D.3.1]{su2025many}. Our study examines implemented visibility and timing policies using numerically scored expressions. The allocation reference changes how a fixed call budget is distributed; the communication comparison holds agent count and rounds fixed. Shared selection permits comparison of candidate archives \citep{li2026dictionaries}, and crossed delivery measures selection dependence with those archives fixed. This design addresses candidate generation and selection in one observable search instance, without explaining performance differences across papers with different tasks and models. Connectivity, timing, and history handling still jointly define each communication policy.

The performance comparisons leave the communication mechanism unresolved. Establishing how a peer message changes search would require linking exposure to subsequent candidate changes and their quality, with an appropriate comparison for that exposure. The present results motivate that test; they do not complete it.

Holding the generated archives fixed, the measured gap between architectures changed with the selection rule: in the Qwen experiment, Complete minus Independent was +0.10 under archive-wide selection and -0.75 under final-round selection, and 9 of 20 prespecified interactions survived Holm correction. A reported ranking is therefore not identified by the architecture alone.

Three reporting practices follow from these comparisons. First, state the selection rule alongside the architecture. Second, report the generated archive and the returned expression as separate endpoints, because a team can produce a better candidate than its own rule returns and a single final score cannot separate those cases. Third, when comparing architectures, either hold the selector fixed across all of them or cross architecture with selector, since architecture-specific rules confound generation with selection. These practices do not depend on the present testbed. They require only that candidate generation be recorded before selection.

Future work should separate topology from update timing, for example by testing parallel scheduling under the same Ring adjacency; test peer messages that include evaluations of candidate expressions; extend the comparison to stronger backbones and broader task families; and repeat conditions across independent samples to quantify run-to-run variation.

None of this implies that multi-agent organization should be avoided. Distributing a fixed call budget across more independent trajectories produced the largest measured gains reported here. In the $N=5$ communication comparison, Star and Complete did not provide clear evidence of an additional gain over Independent, whereas the implemented Ring policy reduced performance under the common archive rule. Whether exchanging partial results among trajectories adds anything beyond the distribution itself remains open.

%% file: sections/appendices.tex
\clearpage\FloatBarrier\setcounter{table}{0}\setcounter{figure}{0}
\section{Experimental environment, prompts, and communication}\label{app:A}

\subsection{Tasks and the information available to each scientist}

We use symbolic regression as a controlled instance of candidate-hypothesis generation and exchange. The Qwen analysis includes 100 tasks: manifest indices 0--49 from LSR-Transform and 111--160 from LSR-Synth. The Terra analysis uses the common 41-task Transform intersection, indices 0--40. Each scientist receives the target name, variable names and descriptions, five training examples, its own previous responses, and the peer responses permitted by the communication policy. The reference expression and held-out arrays are not shown to the scientists. This is therefore an experiment on expression proposals and their exchange under restricted task information; it does not reproduce every stage of scientific inquiry.

The evaluation arrays are pregenerated arrays in the official HDF5 file. The held-out ID test array is disjoint from the training array and is not a split constructed during these experiments. Among the selected Transform tasks, training arrays contain 25,006--80,000 rows and held-out ID arrays contain 6,251--20,000 rows. Forty-three Transform tasks have 80,000 training rows and 20,000 held-out ID rows. Each selected Synth task has 4,000 training rows, 500 held-out ID rows, and 500 OOD rows. The selected Transform tasks have no OOD split. Five training rows enter each prompt; the evaluation procedure uses the pregenerated evaluation arrays rather than these five displayed examples.

\subsection{Invocation schedules and model settings}

All reported fixed-budget conditions allocate 50 scheduled model calls per problem. The independent-allocation comparison changes the number of trajectories and the number of calls along each trajectory together. The communication comparison holds these counts at five agents and ten rounds.

\begingroup\tablefont
\setlength{\tabcolsep}{3.0pt}
\begin{longtable}{@{}>{\raggedright\arraybackslash}p{110.049pt}>{\raggedright\arraybackslash}p{83.486pt}>{\raggedright\arraybackslash}p{87.280pt}>{\raggedright\arraybackslash}p{98.665pt}@{}}
\toprule
\textbf{Condition} & \textbf{Agents or trajectories} & \textbf{Rounds per trajectory} & \textbf{Scheduled calls per problem} \\
\midrule
\endfirsthead
\toprule
\textbf{Condition} & \textbf{Agents or trajectories} & \textbf{Rounds per trajectory} & \textbf{Scheduled calls per problem} \\
\midrule
\endhead
\midrule\multicolumn{4}{r}{\emph{Continued on next page}}\\\endfoot
\bottomrule\endlastfoot
Matched Solo & 1 & 50 & 50 \\
Independent, N=2 & 2 & 25 & 50 \\
Independent, N=5 & 5 & 10 & 50 \\
Independent, N=10 & 10 & 5 & 50 \\
Ring & 5 & 10 & 50 \\
Star & 5 & 10 & 50 \\
Complete & 5 & 10 & 50 \\
\end{longtable}\endgroup

\begingroup\tablefont
\setlength{\tabcolsep}{3.0pt}
\begin{longtable}{@{}>{\raggedright\arraybackslash}p{77.096pt}>{\raggedright\arraybackslash}p{154.192pt}>{\raggedright\arraybackslash}p{154.192pt}@{}}
\toprule
\textbf{Setting} & \textbf{Qwen} & \textbf{Terra} \\
\midrule
\endfirsthead
\toprule
\textbf{Setting} & \textbf{Qwen} & \textbf{Terra} \\
\midrule
\endhead
\midrule\multicolumn{3}{r}{\emph{Continued on next page}}\\\endfoot
\bottomrule\endlastfoot
Model identifier & \nolinkurl{qwen2.5:7b-instruct} & \nolinkurl{gpt-5.6-terra} \\
Service & Local Ollama, OpenAI-compatible endpoint & Official OpenAI endpoint \\
Temperature & 0.70 & 1.00 \\
Output limit & \nolinkurl{max_tokens=512} & \nolinkurl{max_completion_tokens=2048} \\
Reasoning setting & No \nolinkurl{reasoning_effort} field & \nolinkurl{reasoning_effort=none} \\
Displayed training examples & 5 & 5 \\
\end{longtable}\endgroup

The model settings differ between the two layers. We report comparisons within each layer and do not treat their effect sizes as a controlled model-by-policy comparison. A fixed number of calls also does not establish equal token use or equal runtime.

\subsection{Visibility and update timing}

The communication policies specify both which peer responses are visible and when those responses become available. Self-history contains the scientist's own earlier responses; peer text does not accumulate in that self-history. The software constructs each visibility bundle. Scientists do not choose their neighbors or the messages included in that bundle.

\begingroup\tablefont
\setlength{\tabcolsep}{3.0pt}
\begin{longtable}{@{}>{\raggedright\arraybackslash}p{77.096pt}>{\raggedright\arraybackslash}p{154.192pt}>{\raggedright\arraybackslash}p{154.192pt}@{}}
\toprule
\textbf{Policy} & \textbf{Visible peer responses} & \textbf{Update timing} \\
\midrule
\endfirsthead
\toprule
\textbf{Policy} & \textbf{Visible peer responses} & \textbf{Update timing} \\
\midrule
\endhead
\midrule\multicolumn{3}{r}{\emph{Continued on next page}}\\\endfoot
\bottomrule\endlastfoot
Solo & None & One trajectory receives its own previous responses. \\
Independent & None & Each trajectory receives its own previous responses. \\
Ring & The latest response from the predecessor & Agents run in a rotating serial order. A prompt uses the latest predecessor response available at that call. \\
Star & Leaves receive the latest hub response; the hub receives the current leaf responses & Each round has a leaf phase followed by a hub phase. \\
Complete & Previous-round responses from all four peers & Each round uses a fixed previous-round snapshot, preventing access to same-round peer updates. \\
\end{longtable}\endgroup

The implementation makes synchronous API calls sequentially, including conditions whose information schedule is logically parallel. Snapshot construction prevents same-round leakage in Independent and Complete. Consequently, physical call order and information-update order should be distinguished. Ring, Star, and Complete are implemented communication policies, with differences in visibility and timing; the experiment does not isolate graph connectivity while holding every scheduling detail fixed.

The prompt identifies the recipient as a scientist. Its self-history is labeled \texttt{Your\ previous\ responses}; visible peers are anonymized as \texttt{colleague\ 1}, \texttt{colleague\ 2}, and so on. The prompt does not disclose the scientist's numerical identity, graph topology, visibility rule, or hidden peers. This avoids interpreting an explicit topology label as the experimental treatment.

\subsection{Recovered prompt template}

Each scheduled call transports one Chat Completions user message. Braced fields below are populated with task data and the permitted response histories. The raw prompt and output are persisted for every message.

\begin{Verbatim}[breaklines=true,breakanywhere=true,fontsize=\small,frame=single]
You are a scientist working on a symbolic-regression problem.

Task: infer a compact Python expression for {target} using only these variables:
{variable_names}

Variable meanings:
{variable_descriptions}

Training samples are CSV rows in this column order:
{five_training_rows}

[previous discussion]
Your previous responses:
{self_history_or_none}

Previous responses from colleagues:
{visible_peer_history_or_none}

Rules:
1. Write a concise scientific discussion: your hypothesis, how it relates to previous discussion if any, and key assumptions. Be brief.
2. Then write exactly one final machine-readable line:
EQUATION: <python expression using only {variable_names}>
3. Use ASCII Python expression syntax: +, -, *, /, **, sin, cos, exp, log, sqrt, pi.
4. Use only these variables: {variable_names}.
5. Do not use LaTeX.
\end{Verbatim}

The exact runtime \texttt{runner.py} hash for the July Qwen runs is unrecoverable. Persisted prompts and first-call prompt equality provide the evidence for prompt comparability; a recorded repository commit is not a substitute for the unrecovered runtime hash.

\subsection{Implementation and archived public rationales}

The reported experiments use a custom deterministic Python runner. The scheduler determines the permitted message identifiers before each model call; the runner constructs the prompt and directly invokes an OpenAI-compatible Chat Completions endpoint.

The communication treatment is access to peers' \textbf{public rationales}. Text diagnostics measure stated-reasoning diversity, not private chain-of-thought diversity; this operational definition applies to both model layers. Each output is archived before normalization. The archive records the exact prompt, raw output, equation candidate, training NMSE, message identifier, permitted self and peer identifiers, model settings, token counts, and elapsed time. Training fit is used for candidate scoring and selection; automatic NMSE feedback is not injected into the discussion in these frozen experiments.

\clearpage\FloatBarrier\setcounter{table}{0}\setcounter{figure}{0}
\section{Selection, scoring, and statistical conventions}\label{app:B}

\subsection{Candidate validity and delivery rules}

A valid candidate has a nonempty frozen canonical skeleton and finite training NMSE. The following rules select from the candidates already generated for a particular problem and condition. They do not generate additional candidates. Rules may use training-fit statistics; none accesses held-out targets, OOD targets, or the reference expression.

\begingroup\tablefont
\setlength{\tabcolsep}{3.0pt}
\begin{longtable}{@{}>{\raggedright\arraybackslash}p{77.096pt}>{\raggedright\arraybackslash}p{154.192pt}>{\raggedright\arraybackslash}p{154.192pt}@{}}
\toprule
\textbf{Rule} & \textbf{Pool and selection objective} & \textbf{Tie handling} \\
\midrule
\endfirsthead
\toprule
\textbf{Rule} & \textbf{Pool and selection objective} & \textbf{Tie handling} \\
\midrule
\endhead
\midrule\multicolumn{3}{r}{\emph{Continued on next page}}\\\endfoot
\bottomrule\endlastfoot
d1: archive minimum training error & All valid messages across agents and rounds; minimum finite training NMSE & Lexicographically smallest \nolinkurl{message_id} \\
d2: final-round minimum training error & Valid messages in the maximum \nolinkurl{round_idx}; minimum finite training NMSE & Earliest message in archived message order \\
d3: final-round plurality & Valid messages in the maximum \nolinkurl{round_idx}, grouped by frozen canonical skeleton; largest group & Choose the minimum-training-NMSE representative within each tied group, using archive order for equal errors; choose among tied-group representatives by the same criteria \\
d4: global plurality & All valid messages grouped by frozen canonical skeleton; largest group & Same procedure as d3 \\
d5: last message & The final archived message, if valid & No tie \\
d6: complexity penalty & All valid messages parseable by SymPy \citep{meurer2017sympy}; minimum training NMSE plus normalized complexity penalty & Earliest message in archived message order \\
\end{longtable}\endgroup

For d6, the objective for candidate \(c\) is

\[
\operatorname{NMSE}_{\mathrm{train}}(c)
+0.05\,\frac{\operatorname{nodes}(c)}{\operatorname{nodes}_{\max}},
\]

where \texttt{nodes} counts the nodes in a SymPy preorder traversal and the maximum is computed within that problem-condition archive. Every rule returns no selection when its eligible pool is empty; there is no fallback. The last-message rule is a stress test of delivery rather than a recommendation for practical selection.

The six delivery rules belong to the specified reporting family. The decision to emphasize d2, d3, and d4 in the narrative is exploratory. Specifying the rules and choosing which results to feature are separate decisions.

\subsection{Outcome definition and missing selections}

For finite prediction-target pairs, we calculate

\[
\operatorname{NMSE}
=\frac{\frac{1}{n}\sum_i(\hat y_i-y_i)^2}
{\operatorname{Var}_i(y_i)},
\qquad
Q=-\log_{10}(\operatorname{NMSE}+10^{-10}).
\]

Variance is the population variance, using \texttt{ddof=0}. If the variance is nonpositive, the denominator becomes the mean of \(y_i^2\), with a further fallback to 1 if that mean is zero. For nondegenerate targets, the constant-mean predictor has NMSE equal to 1 and \(Q\) approximately zero. The frozen logarithmic stabilizer maps zero NMSE to \(Q=10\); it was not tuned to the results. Higher \(Q\) indicates lower NMSE.

For each selector, a condition-problem cell without a valid selected output is unavailable. Simple-effect comparisons use common problem indices with finite held-out ID \(Q\) in both conditions. Delivery interactions additionally require the scores needed for both delivery rules. The CSV tables retain the exact problem-index lists and valid-pair counts. Qwen simple-effect counts range from 96 to 100 under d1--d5 and equal 99 under d6. Missing selections are not replaced with successful outputs or imputed scores. The reported d1 primary layers have no selector failure.

\subsection{Estimates, confidence intervals, and multiplicity}

The primary effect estimate is the mean paired difference in held-out ID \(Q\). The numerical authority specifies 95\% percentile confidence intervals from 10,000 bootstrap resamples of problems with seed \texttt{20260904}. For contrasts, a resampled problem contributes both condition scores before differencing; for absolute endpoints, resampling applies to problem-level \(Q\). These intervals describe uncertainty across sampled problems and do not constitute repeated-run estimates of model-sampling variability.

We also report medians, wins/losses/ties, signed-rank rank-biserial correlations, and paired Wilcoxon tests computed with SciPy \citep{virtanen2020scipy}. These summaries do not estimate the same quantity as the mean and its bootstrap interval. In particular, a Wilcoxon result and a mean interval can give different indications without either being omitted. Rank-biserial correlations are signed-rank quantities and are not reconstructed from wins/losses/ties.

\begingroup\tablefont
\setlength{\tabcolsep}{3.0pt}
\begin{longtable}{@{}>{\raggedright\arraybackslash}p{77.096pt}>{\raggedright\arraybackslash}p{154.192pt}>{\raggedright\arraybackslash}p{154.192pt}@{}}
\toprule
\textbf{Analysis family} & \textbf{Comparisons in the family} & \textbf{Correction} \\
\midrule
\endfirsthead
\toprule
\textbf{Analysis family} & \textbf{Comparisons in the family} & \textbf{Correction} \\
\midrule
\endhead
\midrule\multicolumn{3}{r}{\emph{Continued on next page}}\\\endfoot
\bottomrule\endlastfoot
Qwen allocation & 1 to 2, 2 to 5, 5 to 10, 1 to 5, and 1 to 10 & Holm across 5 tests \\
Pooled Qwen communication & Ring, Star, and Complete minus Independent under d1 & Holm across 3 exploratory tests \\
Within-stratum allocation & Independent minus matched Solo & One-test family; Holm value equals raw value \\
Within-stratum communication & Ring, Star, and Complete minus Independent & Holm across 3 tests within each stratum/layer \\
Within-stratum policy comparisons & The three pairwise comparisons among Ring, Star, and Complete & Holm across 3 tests within each stratum/layer \\
Delivery simple effects & Each architecture minus Independent under each delivery rule & Exported simple-effect $p$ values are raw \\
Delivery interactions & Four architectures relative to Independent, crossed with d2–d6 relative to d1 & Holm across 20 tests separately for Qwen and Terra \\
\end{longtable}\endgroup

For architecture \(G\), the per-problem delivery interaction is

\[
\bigl(Q_{G,d_k}-Q_{\mathrm{Independent},d_k}\bigr)
-\bigl(Q_{G,d_1}-Q_{\mathrm{Independent},d_1}\bigr).
\]

The interaction table evaluates this difference on its common valid problem set. It should not be reconstructed by subtracting separately reported simple-effect means when their problem sets differ. The tables contain 24 simple effects and 20 interactions for each model layer. Qwen has 9 of 20 Holm-significant interactions and Terra has 0 of 20. The Terra subset had limited sensitivity; its estimates were too imprecise to distinguish small interactions from no interaction.

\clearpage\FloatBarrier\setcounter{table}{0}\setcounter{figure}{0}
\section{Data splits and evaluation diagnostics}\label{app:C}

The common-selector ID comparisons use the same pregenerated arrays under the training-fit selector. The data snapshot is derived from LLM-SRBench \citep{shojaee2025srbench}. The published benchmark reports 239 tasks (111 Transform and 128 Synth), whereas the local HDF5 snapshot used here contains 240 groups (111 Transform and 129 Synth). The 100-task Qwen cohort is a fixed prefix within each selected task family, not a random draw from either full collection; its inference population is therefore the tested cohort. The OOD endpoint is restricted to the 50 selected Synth tasks; the Transform tasks have no OOD array. OOD results are secondary to the amended full-cohort held-out ID endpoint.

The figures below retain the frozen analysis definitions. Distribution plots describe between-problem variation; they do not provide replication over independent model-sampling seeds. Thresholds on NMSE quantify numerical prediction accuracy and must not be called exact symbolic recovery.

\begin{figure}[htbp]
\centering
\includegraphics[width=0.98\linewidth]{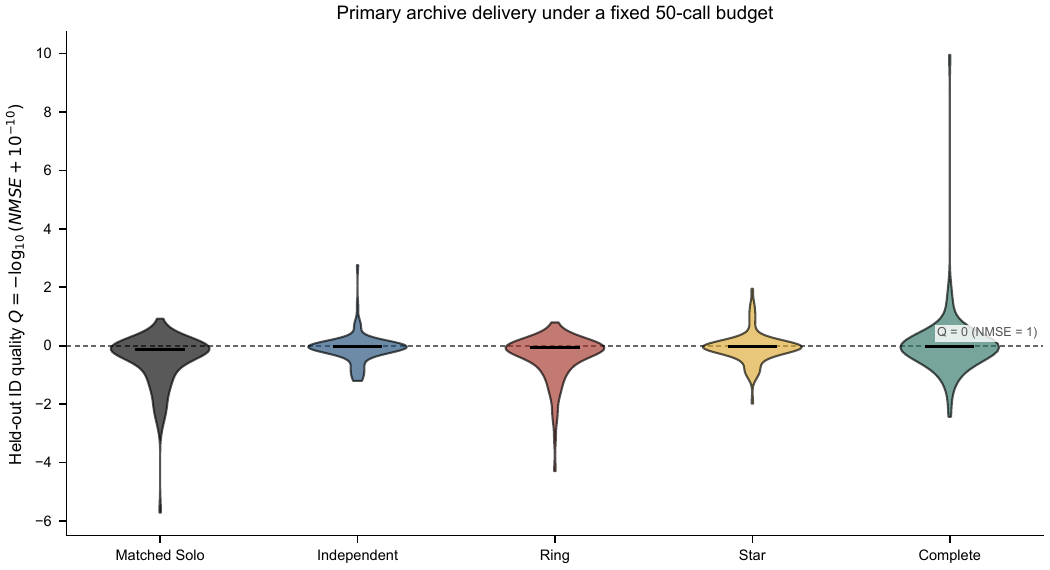}
\caption{Frozen common-rule held-out ID score distributions across the 100 Qwen tasks. Q=0 corresponds approximately to NMSE=1.}
\label{fig:id-distribution}
\end{figure}

\begin{figure}[htbp]
\centering
\includegraphics[width=0.98\linewidth]{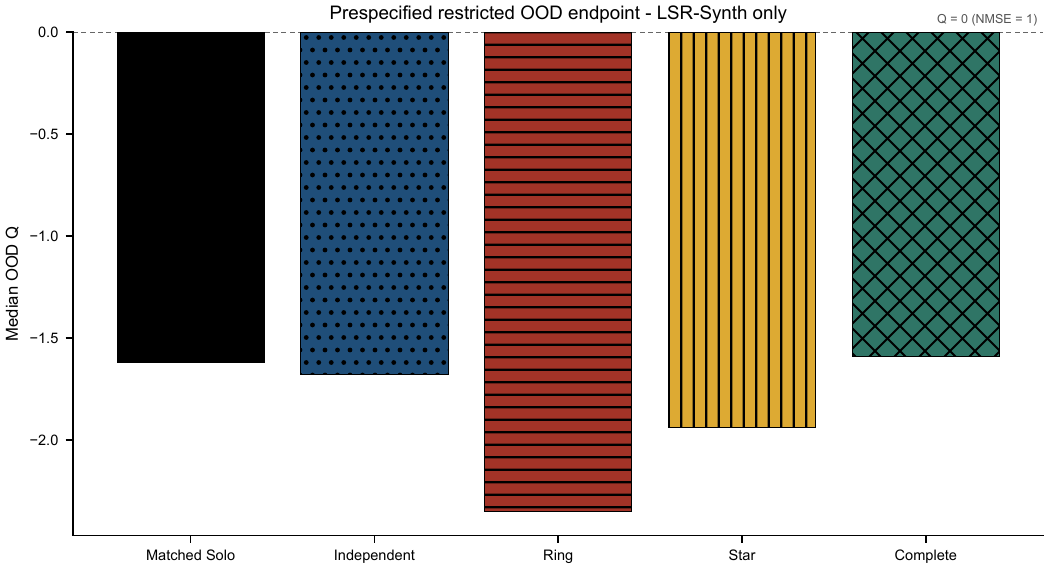}
\caption{Restricted OOD sensitivity on the selected Synth tasks only. These tasks have a pregenerated OOD array; the Transform tasks do not. Q=0 corresponds approximately to NMSE=1.}
\label{fig:restricted-ood}
\end{figure}

\clearpage\FloatBarrier\setcounter{table}{0}\setcounter{figure}{0}
\section{Delivery effects, interactions, and sensitivity}\label{app:D}

\subsection{Reading the delivery tables}

Tables D1 and D2 report all simple effects relative to Independent under the same rule. Tables D3 and D4 report per-problem changes in those gaps relative to d1. Simple-effect p values are raw; interaction p values are also adjusted across the 20 contrasts within each model layer. The 9/20 Qwen and 0/20 Terra counts refer to Holm-adjusted signed-rank tests. The Terra estimates remain too imprecise to distinguish small interactions from no interaction.

Each row uses its own common finite-score intersection. Qwen valid pairs range from 96 to 100. Terra has 41 pairs for d1--d5 and 40 for d6. An interaction requires the four component scores on the same problems; subtracting means from different cohorts would give a different quantity. Exact row-level problem identifiers are retained in the accompanying authority CSVs, identified in the source register below. The rules do not adapt candidate generation, so these are selection comparisons on observed archives.

\begingroup\tablefont
\setlength{\tabcolsep}{2.0pt}
\begin{longtable}{@{}>{\raggedright\arraybackslash}p{48.552pt}>{\raggedright\arraybackslash}p{20.541pt}>{\raggedright\arraybackslash}p{16.807pt}>{\raggedright\arraybackslash}p{44.818pt}>{\raggedright\arraybackslash}p{130.718pt}>{\raggedright\arraybackslash}p{50.420pt}>{\raggedright\arraybackslash}p{61.624pt}@{}}
\caption{Qwen delivery simple effects.}\\
\toprule
\textbf{Architecture} & \textbf{Rule} & \textbf{n} & \textbf{W/L/T} & \textbf{Mean delta Q [95\% CI]} & \textbf{Median} & \textbf{Raw p} \\
\midrule
\endfirsthead
\toprule
\textbf{Architecture} & \textbf{Rule} & \textbf{n} & \textbf{W/L/T} & \textbf{Mean delta Q [95\% CI]} & \textbf{Median} & \textbf{Raw p} \\
\midrule
\endhead
\midrule\multicolumn{7}{r}{\emph{Continued on next page}}\\\endfoot
\bottomrule\endlastfoot
Complete & d1 & 100 & 53/42/5 & 0.10 [-0.08, 0.34] & $1.73\!\times\!10^{-5}$ & 0.18 \\
Complete & d2 & 100 & 29/65/6 & -0.75 [-1.56, -0.25] & \ensuremath{-6.10\times10^{-3}} & $7.27\!\times\!10^{-6}$ \\
Complete & d3 & 100 & 28/65/7 & -0.78 [-1.59, -0.28] & \ensuremath{-9.58\times10^{-3}} & $1.86\!\times\!10^{-6}$ \\
Complete & d4 & 100 & 32/60/8 & -0.59 [-1.40, -0.10] & \ensuremath{-1.38\times10^{-3}} & \ensuremath{3.63\times10^{-3}} \\
Complete & d5 & 98 & 54/39/5 & 1.22 [0.11, 3.12] & $3.93\!\times\!10^{-4}$ & 0.02 \\
Complete & d6 & 99 & 53/39/7 & 0.10 [-0.07, 0.35] & $2.56\!\times\!10^{-4}$ & 0.10 \\
Ring & d1 & 100 & 21/75/4 & -0.33 [-0.49, -0.18] & -0.01 & $1.92\!\times\!10^{-7}$ \\
Ring & d2 & 100 & 24/70/6 & -0.39 [-0.59, -0.21] & \ensuremath{-8.91\times10^{-3}} & $4.75\!\times\!10^{-7}$ \\
Ring & d3 & 100 & 24/69/7 & -0.38 [-0.58, -0.20] & \ensuremath{-8.91\times10^{-3}} & $5.13\!\times\!10^{-7}$ \\
Ring & d4 & 100 & 28/64/8 & -0.26 [-0.44, -0.11] & \ensuremath{-3.66\times10^{-3}} & $2.30\!\times\!10^{-4}$ \\
Ring & d5 & 98 & 49/45/4 & 1.61 [0.18, 4.21] & $2.93\!\times\!10^{-11}$ & 0.17 \\
Ring & d6 & 99 & 25/70/4 & -0.32 [-0.49, -0.18] & -0.01 & $1.08\!\times\!10^{-6}$ \\
Star & d1 & 100 & 43/50/7 & 0.03 [-0.08, 0.13] & $-5.64\!\times\!10^{-7}$ & 0.91 \\
Star & d2 & 99 & 23/71/5 & -0.48 [-0.82, -0.22] & -0.02 & $1.42\!\times\!10^{-7}$ \\
Star & d3 & 99 & 23/70/6 & -0.58 [-0.94, -0.29] & -0.04 & $4.81\!\times\!10^{-8}$ \\
Star & d4 & 100 & 36/56/8 & -0.41 [-0.75, -0.14] & \ensuremath{-1.24\times10^{-3}} & \ensuremath{2.75\times10^{-3}} \\
Star & d5 & 96 & 47/45/4 & 0.17 [-0.26, 0.59] & 0 & 0.25 \\
Star & d6 & 99 & 40/50/9 & 0.03 [-0.08, 0.14] & $-4.75\!\times\!10^{-9}$ & 0.98 \\
Solo & d1 & 100 & 18/78/4 & -0.43 [-0.63, -0.27] & -0.05 & $2.71\!\times\!10^{-10}$ \\
Solo & d2 & 100 & 18/77/5 & -0.51 [-0.72, -0.33] & -0.06 & $1.42\!\times\!10^{-11}$ \\
Solo & d3 & 100 & 19/75/6 & -0.50 [-0.71, -0.33] & -0.05 & $2.58\!\times\!10^{-11}$ \\
Solo & d4 & 100 & 31/64/5 & -0.31 [-0.50, -0.14] & \ensuremath{-9.12\times10^{-3}} & $2.21\!\times\!10^{-5}$ \\
Solo & d5 & 98 & 45/49/4 & 1.49 [\ensuremath{-2.79\times10^{-3}}, 4.13] & $-1.21\!\times\!10^{-7}$ & 0.68 \\
Solo & d6 & 99 & 20/74/5 & -0.43 [-0.63, -0.26] & -0.05 & $1.20\!\times\!10^{-9}$ \\
\end{longtable}\endgroup

\begingroup\tablefont
\setlength{\tabcolsep}{2.0pt}
\begin{longtable}{@{}>{\raggedright\arraybackslash}p{48.552pt}>{\raggedright\arraybackslash}p{20.541pt}>{\raggedright\arraybackslash}p{16.807pt}>{\raggedright\arraybackslash}p{44.818pt}>{\raggedright\arraybackslash}p{130.718pt}>{\raggedright\arraybackslash}p{50.420pt}>{\raggedright\arraybackslash}p{61.624pt}@{}}
\caption{Terra delivery simple effects.}\\
\toprule
\textbf{Architecture} & \textbf{Rule} & \textbf{n} & \textbf{W/L/T} & \textbf{Mean delta Q [95\% CI]} & \textbf{Median} & \textbf{Raw p} \\
\midrule
\endfirsthead
\toprule
\textbf{Architecture} & \textbf{Rule} & \textbf{n} & \textbf{W/L/T} & \textbf{Mean delta Q [95\% CI]} & \textbf{Median} & \textbf{Raw p} \\
\midrule
\endhead
\midrule\multicolumn{7}{r}{\emph{Continued on next page}}\\\endfoot
\bottomrule\endlastfoot
Complete & d1 & 41 & 4/13/24 & -0.36 [-1.43, 0.73] & 0 & 0.08 \\
Complete & d2 & 41 & 3/19/19 & -0.90 [-2.00, 0.10] & 0 & \ensuremath{2.40\times10^{-3}} \\
Complete & d3 & 41 & 4/15/22 & -0.74 [-1.88, 0.30] & 0 & 0.03 \\
Complete & d4 & 41 & 6/12/23 & -0.92 [-2.05, -0.02] & 0 & 0.05 \\
Complete & d5 & 41 & 11/11/19 & \ensuremath{1.08\times10^{-3}} [-0.20, 0.20] & 0 & 0.91 \\
Complete & d6 & 40 & 4/13/23 & -0.37 [-1.49, 0.75] & 0 & 0.08 \\
Ring & d1 & 41 & 2/19/20 & -1.18 [-2.04, -0.48] & 0 & $5.43\!\times\!10^{-4}$ \\
Ring & d2 & 41 & 4/19/18 & -0.89 [-1.87, -0.05] & 0 & \ensuremath{7.44\times10^{-3}} \\
Ring & d3 & 41 & 6/16/19 & -1.03 [-2.06, -0.14] & 0 & 0.02 \\
Ring & d4 & 41 & 7/13/21 & -0.94 [-1.95, -0.05] & 0 & 0.07 \\
Ring & d5 & 41 & 10/16/15 & 0.02 [-1.11, 1.18] & 0 & 0.66 \\
Ring & d6 & 40 & 3/18/19 & -1.21 [-2.08, -0.48] & 0 & $7.02\!\times\!10^{-4}$ \\
Star & d1 & 41 & 3/14/24 & -0.57 [-1.56, 0.41] & 0 & 0.01 \\
Star & d2 & 41 & 4/20/17 & -1.34 [-2.42, -0.44] & 0 & $5.46\!\times\!10^{-4}$ \\
Star & d3 & 41 & 5/16/20 & -1.18 [-2.29, -0.25] & 0 & \ensuremath{9.61\times10^{-3}} \\
Star & d4 & 41 & 6/14/21 & -1.14 [-2.26, -0.21] & 0 & 0.02 \\
Star & d5 & 41 & 10/15/16 & -0.25 [-1.27, 0.71] & 0 & 0.35 \\
Star & d6 & 40 & 3/13/24 & -0.58 [-1.60, 0.42] & 0 & 0.01 \\
Solo & d1 & 41 & 3/18/20 & -1.04 [-2.07, -0.06] & 0 & \ensuremath{1.87\times10^{-3}} \\
Solo & d2 & 41 & 3/20/18 & -1.36 [-2.39, -0.49] & 0 & $1.62\!\times\!10^{-4}$ \\
Solo & d3 & 41 & 4/15/22 & -1.20 [-2.25, -0.31] & 0 & \ensuremath{5.49\times10^{-3}} \\
Solo & d4 & 41 & 7/11/23 & -0.91 [-2.07, 0.14] & 0 & 0.06 \\
Solo & d5 & 41 & 11/12/18 & -0.27 [-1.26, 0.66] & 0 & 0.69 \\
Solo & d6 & 40 & 4/16/20 & -1.06 [-2.10, -0.07] & 0 & \ensuremath{3.18\times10^{-3}} \\
\end{longtable}\endgroup

\begingroup\tablefont
\setlength{\tabcolsep}{2.0pt}
\begin{longtable}{@{}>{\raggedright\arraybackslash}p{44.338pt}>{\raggedright\arraybackslash}p{16.627pt}>{\raggedright\arraybackslash}p{14.779pt}>{\raggedright\arraybackslash}p{36.948pt}>{\raggedright\arraybackslash}p{121.928pt}>{\raggedright\arraybackslash}p{42.490pt}>{\raggedright\arraybackslash}p{46.185pt}>{\raggedright\arraybackslash}p{46.185pt}@{}}
\caption{Qwen d1-referenced delivery interactions.}\\
\toprule
\textbf{Architecture} & \textbf{Rule} & \textbf{n} & \textbf{W/L/T} & \textbf{Mean interaction [95\% CI]} & \textbf{Median interaction} & \textbf{Raw p} & \textbf{Holm p} \\
\midrule
\endfirsthead
\toprule
\textbf{Architecture} & \textbf{Rule} & \textbf{n} & \textbf{W/L/T} & \textbf{Mean interaction [95\% CI]} & \textbf{Median interaction} & \textbf{Raw p} & \textbf{Holm p} \\
\midrule
\endhead
\midrule\multicolumn{8}{r}{\emph{Continued on next page}}\\\endfoot
\bottomrule\endlastfoot
Complete & d2 & 100 & 20/69/11 & -0.84 [-1.85, -0.26] & -0.01 & $6.99\!\times\!10^{-9}$ & $1.05\!\times\!10^{-7}$ \\
Ring & d2 & 100 & 32/30/38 & -0.06 [-0.16, 0.02] & 0 & 0.64 & 1 \\
Star & d2 & 99 & 14/67/18 & -0.51 [-0.84, -0.27] & -0.02 & $8.15\!\times\!10^{-10}$ & $1.39\!\times\!10^{-8}$ \\
Solo & d2 & 100 & 31/26/43 & -0.08 [-0.20, \ensuremath{6.37\times10^{-3}}] & 0 & 0.45 & 1 \\
Complete & d3 & 100 & 21/72/7 & -0.87 [-1.88, -0.28] & -0.03 & $3.84\!\times\!10^{-9}$ & $6.14\!\times\!10^{-8}$ \\
Ring & d3 & 100 & 38/29/33 & -0.05 [-0.15, 0.03] & 0 & 0.80 & 1 \\
Star & d3 & 99 & 17/69/13 & -0.61 [-0.94, -0.34] & -0.02 & $3.47\!\times\!10^{-10}$ & $6.60\!\times\!10^{-9}$ \\
Solo & d3 & 100 & 38/25/37 & -0.07 [-0.18, 0.01] & 0 & 0.93 & 1 \\
Complete & d4 & 100 & 37/54/9 & -0.69 [-1.69, -0.10] & $-3.60\!\times\!10^{-4}$ & \ensuremath{6.62\times10^{-3}} & 0.07 \\
Ring & d4 & 100 & 51/20/29 & 0.06 [-0.02, 0.14] & $6.04\!\times\!10^{-7}$ & \ensuremath{1.28\times10^{-3}} & 0.02 \\
Star & d4 & 100 & 31/58/11 & -0.44 [-0.76, -0.19] & \ensuremath{-2.00\times10^{-3}} & $2.19\!\times\!10^{-4}$ & \ensuremath{2.84\times10^{-3}} \\
Solo & d4 & 100 & 55/14/31 & 0.13 [0.05, 0.21] & $1.54\!\times\!10^{-6}$ & $1.09\!\times\!10^{-5}$ & $1.53\!\times\!10^{-4}$ \\
Complete & d5 & 98 & 56/38/4 & 1.12 [0.12, 2.83] & $6.64\!\times\!10^{-4}$ & 0.03 & 0.28 \\
Ring & d5 & 98 & 78/15/5 & 1.94 [0.50, 4.55] & 0.06 & $2.88\!\times\!10^{-10}$ & $5.77\!\times\!10^{-9}$ \\
Star & d5 & 96 & 50/41/5 & 0.15 [-0.26, 0.54] & $6.34\!\times\!10^{-5}$ & 0.32 & 1 \\
Solo & d5 & 98 & 76/12/10 & 1.93 [0.46, 4.56] & 0.05 & $7.17\!\times\!10^{-10}$ & $1.29\!\times\!10^{-8}$ \\
Complete & d6 & 99 & 17/10/72 & \ensuremath{1.42\times10^{-4}} [\ensuremath{-6.86\times10^{-5}}, \ensuremath{3.76\times10^{-4}}] & 0 & 0.10 & 0.72 \\
Ring & d6 & 99 & 17/11/71 & \ensuremath{-5.42\times10^{-4}} [\ensuremath{-2.11\times10^{-3}}, \ensuremath{3.56\times10^{-4}}] & 0 & 0.08 & 0.67 \\
Star & d6 & 99 & 16/17/66 & \ensuremath{1.35\times10^{-4}} [\ensuremath{-1.57\times10^{-4}}, \ensuremath{4.75\times10^{-4}}] & 0 & 0.65 & 1 \\
Solo & d6 & 99 & 19/8/72 & \ensuremath{3.39\times10^{-4}} [\ensuremath{9.49\times10^{-5}}, \ensuremath{6.62\times10^{-4}}] & 0 & \ensuremath{9.47\times10^{-3}} & 0.09 \\
\end{longtable}\endgroup

\begingroup\tablefont
\setlength{\tabcolsep}{2.0pt}
\begin{longtable}{@{}>{\raggedright\arraybackslash}p{44.338pt}>{\raggedright\arraybackslash}p{16.627pt}>{\raggedright\arraybackslash}p{14.779pt}>{\raggedright\arraybackslash}p{36.948pt}>{\raggedright\arraybackslash}p{121.928pt}>{\raggedright\arraybackslash}p{42.490pt}>{\raggedright\arraybackslash}p{46.185pt}>{\raggedright\arraybackslash}p{46.185pt}@{}}
\caption{Terra d1-referenced delivery interactions.}\\
\toprule
\textbf{Architecture} & \textbf{Rule} & \textbf{n} & \textbf{W/L/T} & \textbf{Mean interaction [95\% CI]} & \textbf{Median interaction} & \textbf{Raw p} & \textbf{Holm p} \\
\midrule
\endfirsthead
\toprule
\textbf{Architecture} & \textbf{Rule} & \textbf{n} & \textbf{W/L/T} & \textbf{Mean interaction [95\% CI]} & \textbf{Median interaction} & \textbf{Raw p} & \textbf{Holm p} \\
\midrule
\endhead
\midrule\multicolumn{8}{r}{\emph{Continued on next page}}\\\endfoot
\bottomrule\endlastfoot
Complete & d2 & 41 & 6/15/20 & -0.54 [-1.51, 0.35] & 0 & 0.05 & 0.89 \\
Ring & d2 & 41 & 8/7/26 & 0.29 [-0.53, 1.08] & 0 & 0.31 & 1 \\
Star & d2 & 41 & 6/18/17 & -0.77 [-1.91, 0.22] & 0 & 0.04 & 0.75 \\
Solo & d2 & 41 & 7/9/25 & -0.32 [-1.40, 0.66] & 0 & 0.68 & 1 \\
Complete & d3 & 41 & 8/13/20 & -0.38 [-1.40, 0.55] & 0 & 0.31 & 1 \\
Ring & d3 & 41 & 10/9/22 & 0.15 [-0.80, 1.06] & 0 & 0.57 & 1 \\
Star & d3 & 41 & 8/14/19 & -0.61 [-1.79, 0.40] & 0 & 0.31 & 1 \\
Solo & d3 & 41 & 11/9/21 & -0.16 [-1.27, 0.86] & 0 & 0.68 & 1 \\
Complete & d4 & 41 & 10/12/19 & -0.56 [-1.69, 0.46] & 0 & 0.47 & 1 \\
Ring & d4 & 41 & 14/5/22 & 0.24 [-0.70, 1.14] & 0 & 0.15 & 1 \\
Star & d4 & 41 & 10/12/19 & -0.58 [-1.74, 0.43] & 0 & 0.49 & 1 \\
Solo & d4 & 41 & 14/6/21 & 0.12 [-0.86, 1.03] & 0 & 0.22 & 1 \\
Complete & d5 & 41 & 14/8/19 & 0.36 [-0.77, 1.46] & 0 & 0.22 & 1 \\
Ring & d5 & 41 & 19/5/17 & 1.20 [-0.02, 2.44] & 0 & \ensuremath{6.64\times10^{-3}} & 0.13 \\
Star & d5 & 41 & 14/12/15 & 0.32 [-1.10, 1.72] & 0 & 0.62 & 1 \\
Solo & d5 & 41 & 16/8/17 & 0.77 [-0.52, 2.09] & 0 & 0.09 & 1 \\
Complete & d6 & 40 & 2/1/37 & \ensuremath{1.48\times10^{-6}} [\ensuremath{-2.04\times10^{-6}}, \ensuremath{6.49\times10^{-6}}] & 0 & 0.59 & 1 \\
Ring & d6 & 40 & 2/2/36 & \ensuremath{-1.65\times10^{-5}} [\ensuremath{-5.03\times10^{-5}}, \ensuremath{5.21\times10^{-7}}] & 0 & 1 & 1 \\
Star & d6 & 40 & 2/2/36 & \ensuremath{-4.52\times10^{-4}} [\ensuremath{-1.33\times10^{-3}}, \ensuremath{3.88\times10^{-7}}] & 0 & 0.47 & 1 \\
Solo & d6 & 40 & 2/2/36 & \ensuremath{-4.58\times10^{-6}} [\ensuremath{-2.04\times10^{-5}}, \ensuremath{6.43\times10^{-6}}] & 0 & 1 & 1 \\
\end{longtable}\endgroup

\subsection{Sign changes and expression grouping}

The source analysis assigns the following classifications. A significant interaction is not sufficient to establish opposite effects on the two sides of a sign change: the simple-effect uncertainty also matters. Complete's d1 mean interval includes zero.

\begingroup\tablefont
\setlength{\tabcolsep}{3.0pt}
\begin{longtable}{@{}>{\raggedright\arraybackslash}p{77.096pt}>{\raggedright\arraybackslash}p{154.192pt}>{\raggedright\arraybackslash}p{154.192pt}@{}}
\toprule
\textbf{Candidate sign change} & \textbf{Source classification} & \textbf{Scope} \\
\midrule
\endfirsthead
\toprule
\textbf{Candidate sign change} & \textbf{Source classification} & \textbf{Scope} \\
\midrule
\endhead
\midrule\multicolumn{3}{r}{\emph{Continued on next page}}\\\endfoot
\bottomrule\endlastfoot
Complete d2 & Point-estimate sign change & The d1 advantage is not established. \\
Complete d3 & Point-estimate sign change & The d1 advantage is not established. \\
Complete d4 & Not supported after correction & The original interaction already fails Holm correction. \\
Star d5 & Not supported after correction & The interaction does not establish a change. \\
Ring d5 & Supported sign-change classification & Confined to the last-message stress test, the weakest rule in this batch. \\
\end{longtable}\endgroup

The alternative functional canonicalization retains the Complete d3 interaction (Holm p = \ensuremath{8.10\times10^{-8}}). Complete d4 fails under both the original grouping (Holm p = 0.07) and the alternative (p = 0.16). The d2 result does not change because its rule does not use skeleton groups. These canonicalization results and the classification table come from the earlier writing package, Part 2.50; the September 5 supplement provides the full original-grouping interactions. No new canonicalization analysis was run for this rewrite.

\begin{figure}[htbp]
\centering
\includegraphics[width=0.98\linewidth]{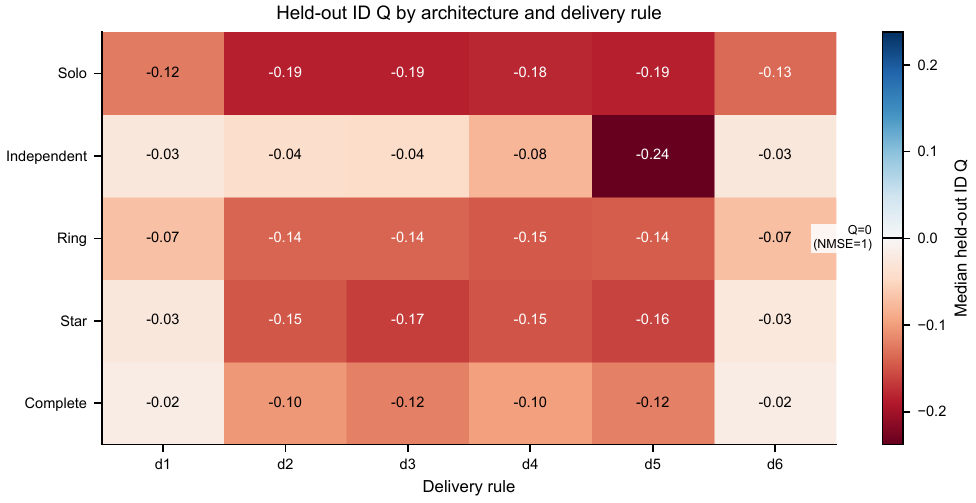}
\caption{Full Qwen architecture by delivery-rule matrix on frozen candidate archives. Cells summarize held-out ID Q; zero corresponds approximately to NMSE=1. Availability depends on the selector.}
\label{fig:delivery-matrix}
\end{figure}

\begin{figure}[htbp]
\centering
\includegraphics[width=0.98\linewidth]{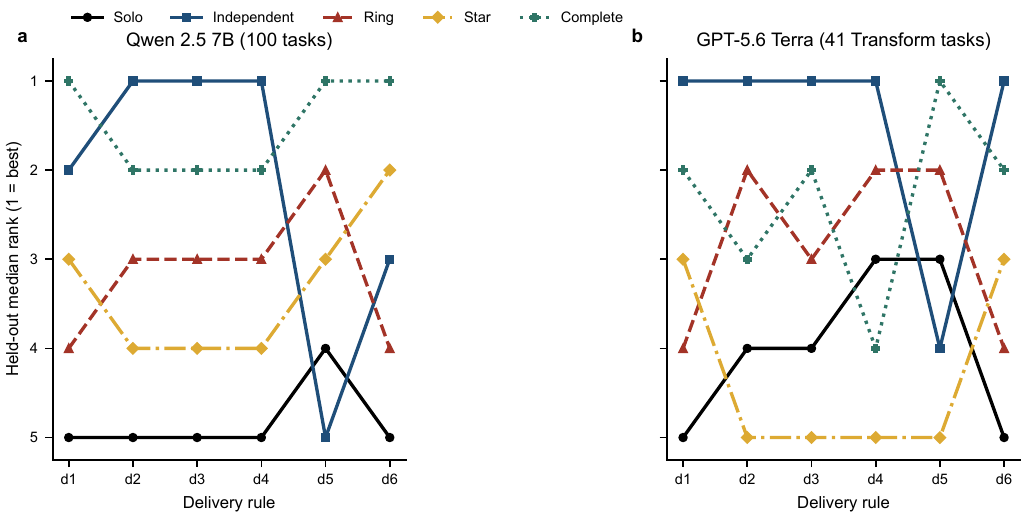}
\caption{Within-layer delivery rankings for Qwen and the Terra Transform subset. Rankings are descriptive; effect magnitudes are not compared across model layers.}
\label{fig:terra-delivery}
\end{figure}

\clearpage\FloatBarrier\setcounter{table}{0}\setcounter{figure}{0}
\section{Allocation and absolute fitting performance}\label{app:E}

Table~\ref{tab:main-contrasts}B gives the complete five-contrast Qwen allocation family. Table E1 adds the absolute endpoints and threshold counts. All conditions use 50 calls per problem and d1 delivery. Intervals concern means; quantiles and threshold counts describe different aspects of the distribution.

\begingroup\tablefont
\setlength{\tabcolsep}{2.0pt}
\begin{longtable}{@{}>{\raggedright\arraybackslash}p{11.084pt}>{\raggedright\arraybackslash}p{121.928pt}>{\raggedright\arraybackslash}p{38.795pt}>{\raggedright\arraybackslash}p{38.795pt}>{\raggedright\arraybackslash}p{40.643pt}>{\raggedright\arraybackslash}p{38.795pt}>{\raggedright\arraybackslash}p{38.795pt}>{\raggedright\arraybackslash}p{40.643pt}@{}}
\caption{Absolute Qwen endpoints.}\\
\toprule
\textbf{N} & \textbf{Mean Q [95\% CI]} & \textbf{Median} & \textbf{Q1} & \textbf{Q3} & \textbf{NMSE < 1} & \textbf{NMSE < 0.10} & \textbf{NMSE < 0.01} \\
\midrule
\endfirsthead
\toprule
\textbf{N} & \textbf{Mean Q [95\% CI]} & \textbf{Median} & \textbf{Q1} & \textbf{Q3} & \textbf{NMSE < 1} & \textbf{NMSE < 0.10} & \textbf{NMSE < 0.01} \\
\midrule
\endhead
\midrule\multicolumn{8}{r}{\emph{Continued on next page}}\\\endfoot
\bottomrule\endlastfoot
1 & -0.53 [-0.71, -0.36] & -0.12 & -0.84 & -0.01 & 11 & 0 & 0 \\
2 & -0.30 [-0.45, -0.15] & -0.06 & -0.45 & \ensuremath{-5.62\times10^{-3}} & 12 & 3 & 0 \\
5 & -0.10 [-0.19, \ensuremath{5.49\times10^{-3}}] & -0.03 & -0.18 & $-3.60\!\times\!10^{-5}$ & 19 & 2 & 1 \\
10 & 0.17 [-0.08, 0.49] & \ensuremath{-1.70\times10^{-3}} & -0.12 & 0.05 & 38 & 4 & 3 \\
\end{longtable}\endgroup

At N=10, 38/100 problems beat constant-mean prediction, while the mean interval {[}-0.08, +0.49{]} includes zero. The adjacent comparisons have 5, 8, and 5 exact ties. Thus, small median gains do not imply unchanged outputs, and strong paired gains do not establish strong absolute fitting performance.

An exploratory lower-tail summary uses the strict threshold Q \textless{} Q10(N=5), whose value is -0.76 in the numerical authority. The counts are 28, 16, 10, and 6 for N=1, 2, 5, and 10. In the communication comparison, the earlier writing package reports 10 Independent, 23 Ring, 11 Star, and 12 Complete cases below this threshold. These descriptive counts are not an independent confirmation of the primary mean comparison.

\begin{figure}[htbp]
\centering
\includegraphics[width=0.98\linewidth]{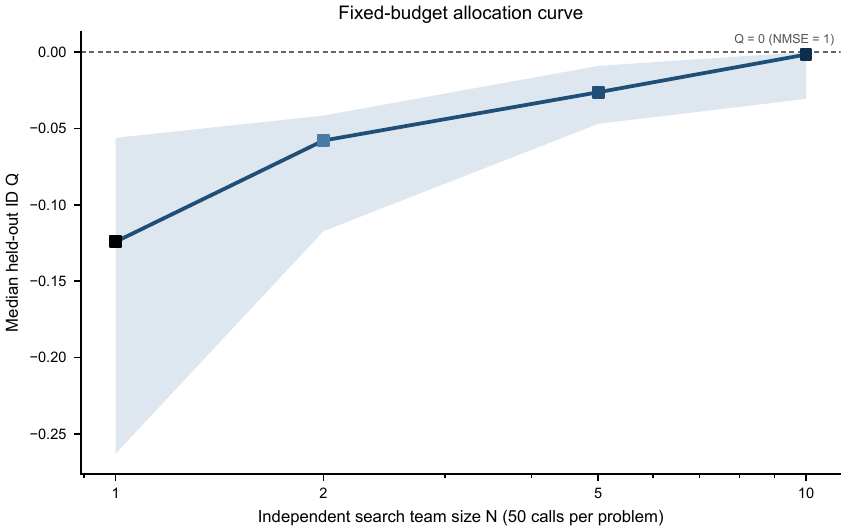}
\caption{Median held-out ID Q under fixed-budget allocation. This descriptive curve complements the mean and confidence intervals in Figure~\ref{fig:3}; Q=0 corresponds approximately to NMSE=1.}
\label{fig:allocation-full}
\end{figure}

\clearpage\FloatBarrier\setcounter{table}{0}\setcounter{figure}{0}
\section{Qwen task-family contrasts}\label{app:F}

The following tables include all six exported contrasts within each task family. The single allocation contrast, the three communication contrasts, and the three within-policy contrasts belong to distinct Holm families; the Star--Complete policy test appears in Appendix~\ref{app:J}'s sensitivity table where provided. The main Transform values retain the historical nonzero-input convention. Raw and adjusted p values are separately labeled. All pooled communication contrasts remain exploratory.

\subsection{LSR-Transform}

\begingroup\tablefont
\setlength{\tabcolsep}{2.0pt}
\begin{longtable}{@{}>{\raggedright\arraybackslash}p{72.049pt}>{\raggedright\arraybackslash}p{12.932pt}>{\raggedright\arraybackslash}p{101.607pt}>{\raggedright\arraybackslash}p{44.338pt}>{\raggedright\arraybackslash}p{36.948pt}>{\raggedright\arraybackslash}p{35.101pt}>{\raggedright\arraybackslash}p{33.253pt}>{\raggedright\arraybackslash}p{33.253pt}@{}}
\toprule
\textbf{Contrast} & \textbf{n} & \textbf{Mean delta Q [95\% CI]} & \textbf{Median} & \textbf{W/L/T} & \textbf{r\_rb} & \textbf{Raw p} & \textbf{Holm p} \\
\midrule
\endfirsthead
\toprule
\textbf{Contrast} & \textbf{n} & \textbf{Mean delta Q [95\% CI]} & \textbf{Median} & \textbf{W/L/T} & \textbf{r\_rb} & \textbf{Raw p} & \textbf{Holm p} \\
\midrule
\endhead
\midrule\multicolumn{8}{r}{\emph{Continued on next page}}\\\endfoot
\bottomrule\endlastfoot
Independent - Matched Solo & 50 & 0.35 [0.16, 0.56] & 0.07 & 34/12/4 & 0.60 & $2.31\!\times\!10^{-4}$ & $2.31\!\times\!10^{-4}$ \\
Ring - Independent & 50 & -0.30 [-0.48, -0.14] & -0.02 & 10/36/4 & -0.64 & $7.84\!\times\!10^{-5}$ & $2.35\!\times\!10^{-4}$ \\
Star - Independent & 50 & 0.11 [-0.03, 0.27] & 0 & 24/19/7 & 0.19 & 0.29 & 0.29 \\
Complete - Independent & 50 & 0.23 [-0.08, 0.70] & $1.73\!\times\!10^{-5}$ & 28/17/5 & 0.28 & 0.11 & 0.21 \\
Ring - Star & 50 & -0.41 [-0.62, -0.24] & -0.02 & 9/36/5 & -0.75 & $2.29\!\times\!10^{-6}$ & $4.57\!\times\!10^{-6}$ \\
Ring - Complete & 50 & -0.53 [-1.00, -0.22] & -0.19 & 7/39/4 & -0.79 & $4.36\!\times\!10^{-7}$ & $1.31\!\times\!10^{-6}$ \\
\end{longtable}\endgroup

\subsection{LSR-Synth}

\begingroup\tablefont
\setlength{\tabcolsep}{2.0pt}
\begin{longtable}{@{}>{\raggedright\arraybackslash}p{72.049pt}>{\raggedright\arraybackslash}p{12.932pt}>{\raggedright\arraybackslash}p{101.607pt}>{\raggedright\arraybackslash}p{44.338pt}>{\raggedright\arraybackslash}p{36.948pt}>{\raggedright\arraybackslash}p{35.101pt}>{\raggedright\arraybackslash}p{33.253pt}>{\raggedright\arraybackslash}p{33.253pt}@{}}
\toprule
\textbf{Contrast} & \textbf{n} & \textbf{Mean delta Q [95\% CI]} & \textbf{Median} & \textbf{W/L/T} & \textbf{r\_rb} & \textbf{Raw p} & \textbf{Holm p} \\
\midrule
\endfirsthead
\toprule
\textbf{Contrast} & \textbf{n} & \textbf{Mean delta Q [95\% CI]} & \textbf{Median} & \textbf{W/L/T} & \textbf{r\_rb} & \textbf{Raw p} & \textbf{Holm p} \\
\midrule
\endhead
\midrule\multicolumn{8}{r}{\emph{Continued on next page}}\\\endfoot
\bottomrule\endlastfoot
Independent - Matched Solo & 50 & 0.52 [0.26, 0.84] & 0.04 & 44/6/0 & 0.85 & $5.80\!\times\!10^{-9}$ & $5.80\!\times\!10^{-9}$ \\
Ring - Independent & 50 & -0.35 [-0.63, -0.12] & -0.01 & 11/39/0 & -0.58 & $2.21\!\times\!10^{-4}$ & $6.63\!\times\!10^{-4}$ \\
Star - Independent & 50 & -0.06 [-0.21, 0.07] & \ensuremath{-2.03\times10^{-3}} & 19/31/0 & -0.21 & 0.20 & 0.40 \\
Complete - Independent & 50 & -0.04 [-0.18, 0.09] & $6.44\!\times\!10^{-5}$ & 25/25/0 & 0.03 & 0.84 & 0.84 \\
Ring - Star & 50 & -0.30 [-0.55, -0.08] & -0.01 & 14/36/0 & -0.50 & \ensuremath{1.78\times10^{-3}} & \ensuremath{3.56\times10^{-3}} \\
Ring - Complete & 50 & -0.32 [-0.61, -0.07] & -0.02 & 12/38/0 & -0.57 & $3.24\!\times\!10^{-4}$ & $9.72\!\times\!10^{-4}$ \\
\end{longtable}\endgroup

\begin{figure}[htbp]
\centering
\includegraphics[width=0.98\linewidth]{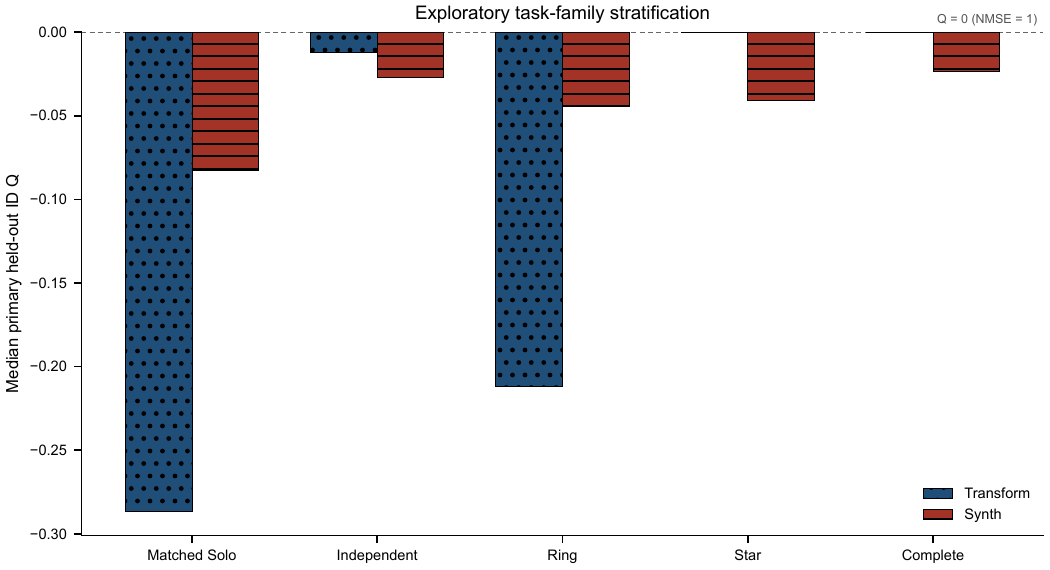}
\caption{Absolute median Qwen held-out ID scores by task family under the common selector. Q=0 corresponds approximately to NMSE=1. The displayed family-specific ordering is descriptive and does not establish an interaction.}
\label{fig:stratified}
\end{figure}

\clearpage\FloatBarrier\setcounter{table}{0}\setcounter{figure}{0}
\section{Exploratory process diagnostics}\label{app:G}

\subsection{Modal-skeleton concentration}

The concentration summary counts cases in which a modal canonical skeleton occupies at least 80\% of the recorded outputs. Problem-level pooling includes all agents; an agent-level summary uses an individual trajectory. The source package reports the following Qwen summaries.

\begingroup\tablefont
\setlength{\tabcolsep}{3.0pt}
\begin{longtable}{@{}>{\raggedright\arraybackslash}p{97.870pt}>{\raggedright\arraybackslash}p{293.610pt}@{}}
\toprule
\textbf{Policy} & \textbf{Problems meeting the concentration threshold} \\
\midrule
\endfirsthead
\toprule
\textbf{Policy} & \textbf{Problems meeting the concentration threshold} \\
\midrule
\endhead
\midrule\multicolumn{2}{r}{\emph{Continued on next page}}\\\endfoot
\bottomrule\endlastfoot
Solo & 81 \\
Ring & 72 \\
Star & 56 \\
Complete & 46 \\
Independent & 0 \\
\end{longtable}\endgroup

The reported agent-level rates are 81\% for Solo and 81.80\% for Independent. These are time-aggregated summaries. They do not establish absorption, the direction of information transfer, or the causal effect of an adopted peer expression. In particular, pooling different repetitive trajectories can yield low problem-level concentration.

\subsection{Within-trajectory improvement}

Solo's best-so-far Q improves by 0.01 between calls 10 and 50. This observation excludes an interpretation in which the allocation comparison establishes that iteration produces no improvement. It does not explain the communication-policy differences. The present manuscript uses these diagnostics to limit process interpretations and does not add a mechanism claim from lexical similarity, symbolic distance, or a temporal gate.

\begin{figure}[htbp]
\centering
\includegraphics[width=0.98\linewidth]{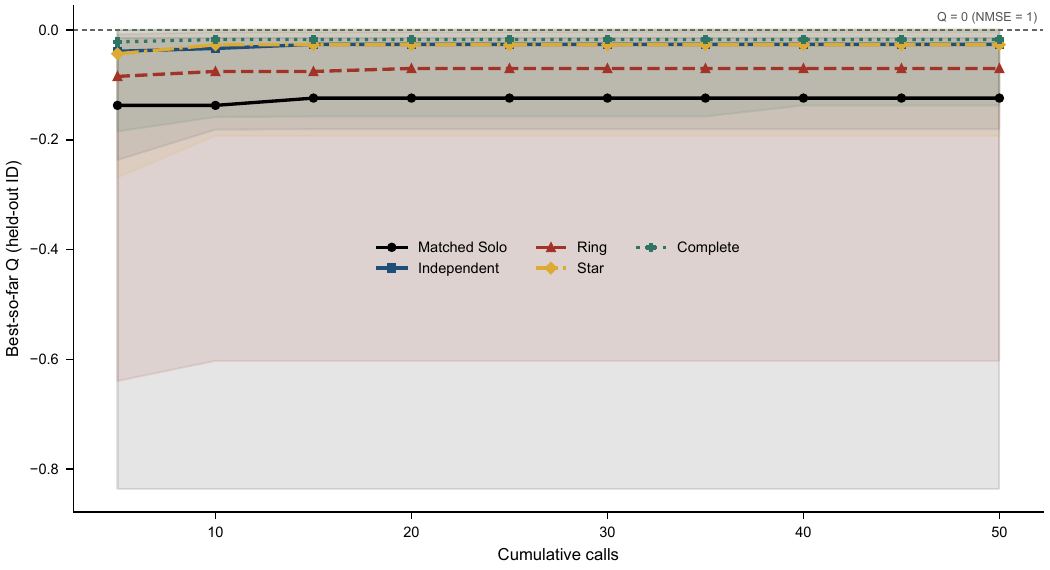}
\caption{Exploratory archived best-so-far trajectories versus cumulative calls. Candidate selection and the evaluation split follow the plotted analysis. These archive diagnostics do not identify causal peer influence.}
\label{fig:trajectories}
\end{figure}

\begin{figure}[htbp]
\centering
\includegraphics[width=0.98\linewidth]{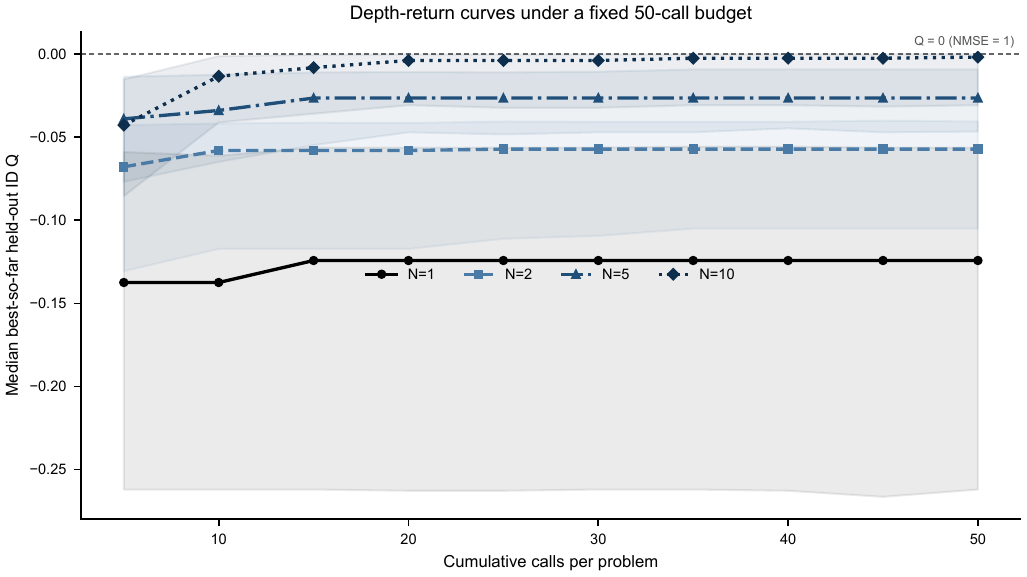}
\caption{Exploratory depth-return curves under the fixed-budget allocations. A flat aggregate curve does not imply that every individual trajectory is unchanged. Q=0 corresponds approximately to NMSE=1.}
\label{fig:depth}
\end{figure}

\clearpage
\subsection{A post-hoc search for a diversity mechanism}\label{app:G3}

After the primary comparisons were complete, we tested whether the communication results could be described by a reduction in the number of distinct candidates generated. The account was that delivery selects an extremum over the archive, so the value of additional agents depends on how different their candidates are. This analysis is post-hoc and exploratory. It belongs to neither the primary communication family nor the prespecified delivery family.

\textbf{Definition.} For each problem and condition, $D$ counts the distinct frozen canonical skeletons among candidates with a nonempty skeleton and finite stored training NMSE. Median $D$ was 2 for $N=1$ and Solo, 3 for $N=2$, 8.50 for $N=5$ and Independent, 15.50 for $N=10$, 2 for Ring, 5 for Star, and 7 for Complete.

\textbf{First test and its stopping rule.} We specified in advance that the analysis would proceed only if the paired within-problem correlation between $\Delta D$ and $\Delta Q$ was positive with raw $p < 0.05$ for all three communication contrasts. Across 100 problems the Spearman correlations were $-0.02$ for Ring ($p = 0.85$), $+0.26$ for Star ($p = 0.01$), and $+0.08$ for Complete ($p = 0.46$); the Solo control was $+0.01$ ($p = 0.89$). The rule was triggered and the downstream analyses were not run.

\textbf{Deviation.} We subsequently ran the stratified test that the stopping rule had blocked. This is a deviation, decided after the first results were known. The reason is that the stopping rule used a pooled correlation, whereas the stratified prediction was that the correlation would have opposite signs depending on whether the modal skeleton was also the best-scoring one in the archive. Under that prediction the pooled correlation is expected to be near zero, so the pooled test cannot distinguish the prediction holding from no effect in either stratum. The stratum assignment rule and the interaction criterion were fixed before the stratified statistics were computed.

\textbf{Stratified result.} Assigning each paired observation by the stratum of the communicating condition, the difference in Spearman correlation between the modal-is-best and modal-is-not-best strata was $+0.13$ for Ring (95\% CI $[-0.30, 0.55]$), $-0.15$ for Star ($[-0.66, 0.33]$), and $+0.03$ for Complete ($[-0.46, 0.53]$). No interaction was detected. Ring's stratum-specific correlations were in the predicted direction, $+0.02$ and $-0.11$, but the interval spans zero. In a sensitivity analysis assigning observations by the stratum of Independent, the Star interval was $[-0.84, -0.02]$, which excludes zero in the direction opposite to the prediction; the other intervals spanned zero. No frozen version of the analysis supported the account.

\textbf{Status.} The diversity account is not supported by these data. Neither is it falsified by the strict criterion we recorded, which required both strata to be significantly negative. The counts in Section 3.20 remain as descriptions of what each condition generated. The archives, the export used for these statistics, and the analysis scripts are retained in the source register.

\clearpage\FloatBarrier\setcounter{table}{0}\setcounter{figure}{0}
\section{Extended exploratory archive diagnostics}\label{app:H}

These panels are post-hoc exploratory and belong to neither the primary communication family nor the prespecified delivery-interaction family. They are included to document available observations, with their original analysis units and limitations. Preparing the panels in this Appendix H section during manuscript assembly introduced no additional experiments or statistical tests; the separate later diversity-analysis timeline is recorded in Appendix~\ref{app:G3} and Appendix~\ref{app:J}.

\subsection{Function and symbolic spaces}

Symbolic distance uses the authorized operator/variable multiset-Jaccard fallback. It is not tree edit distance. The reported symbolic space may identify numerically different expressions, so concentration in that space is insufficient evidence of agreement in function values. Normalized dispersion ratios are undefined where the baseline or denominator is zero; those units are excluded from their aggregates. The single-problem illustration fits its PCA and MDS to that problem's point set. Coordinates have no across-problem meaning, and the illustration does not establish a population mechanism.

\subsection{Lexical overlap and its limits}

Equation-token-stripped lexical overlap removes variable names, target names, equation tokens, and common mathematical functions before calculating Jaccard overlap. The null pool remains same-problem nonrecipient candidates. This helps address literal equation copying and some shared-topic vocabulary, but does not identify whether a response was causally influenced by a particular colleague. Blinded human validation of the automatic public-rationale coding remains pending.

\begingroup\tablefont
\setlength{\tabcolsep}{3.0pt}
\begin{longtable}{@{}>{\raggedright\arraybackslash}p{77.096pt}>{\raggedright\arraybackslash}p{154.192pt}>{\raggedright\arraybackslash}p{154.192pt}@{}}
\toprule
\textbf{Policy} & \textbf{Original excess overlap} & \textbf{After token stripping} \\
\midrule
\endfirsthead
\toprule
\textbf{Policy} & \textbf{Original excess overlap} & \textbf{After token stripping} \\
\midrule
\endhead
\midrule\multicolumn{3}{r}{\emph{Continued on next page}}\\\endfoot
\bottomrule\endlastfoot
Ring & 0.15 & 0.15 \\
Star & 0.11 & 0.12 \\
Complete & 0.02 & 0.03 \\
\end{longtable}\endgroup

These descriptive values are not added to the core estimand family. A causal influence interpretation would require an exposure comparison that separates common task responses from source-specific information transfer.

\begin{figure}[htbp]
\centering
\includegraphics[width=0.98\linewidth]{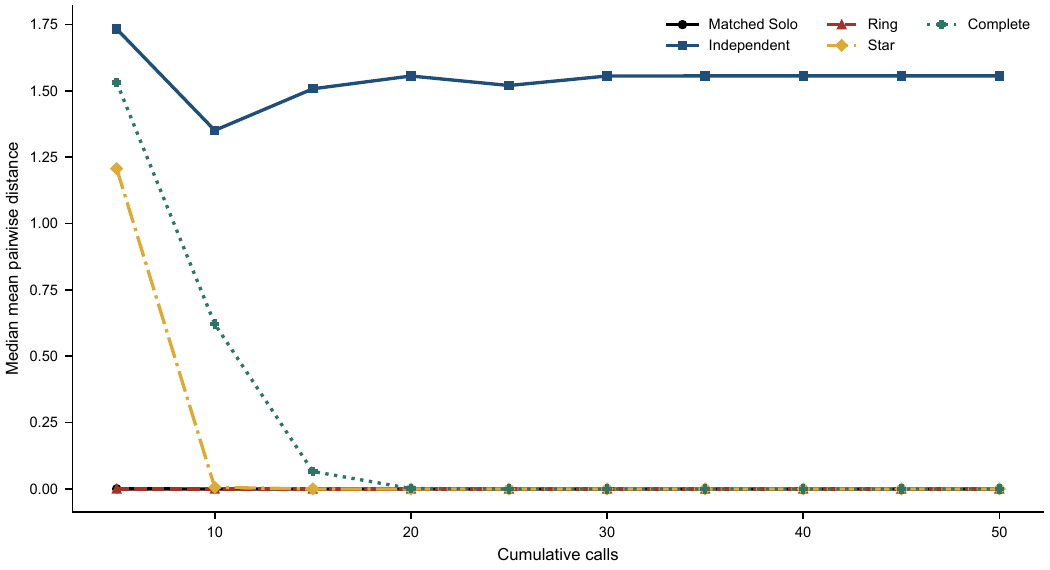}
\caption{Exploratory function-space dispersion over checkpoints. Between-problem summaries preserve the analysis-specific valid-unit denominators.}
\label{fig:function-dispersion}
\end{figure}

\begin{figure}[htbp]
\centering
\includegraphics[width=0.98\linewidth]{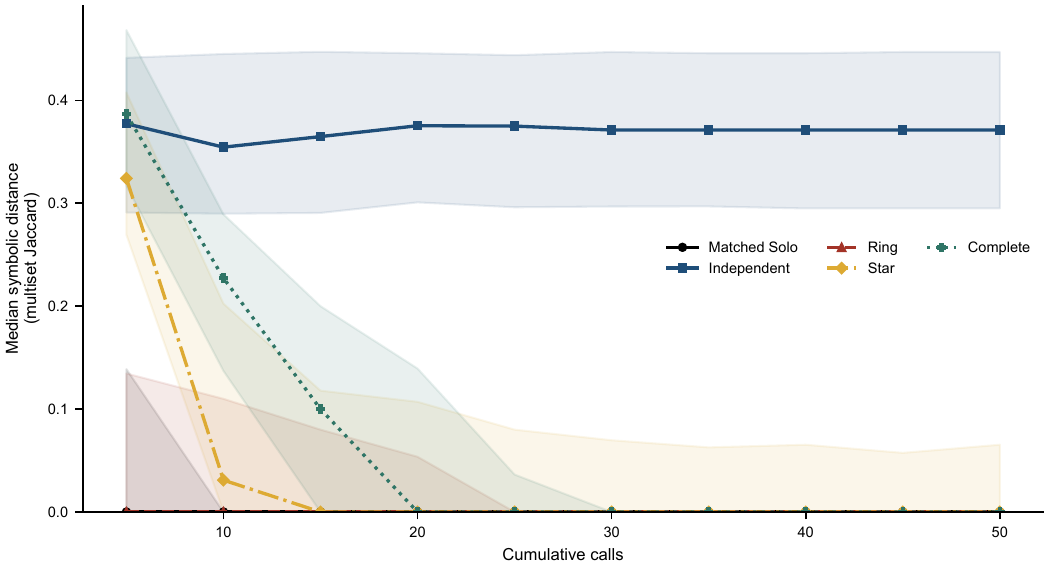}
\caption{Exploratory symbolic dispersion using operator/variable multiset Jaccard, not tree edit distance.}
\label{fig:symbolic-dispersion}
\end{figure}

\begin{figure}[htbp]
\centering
\includegraphics[width=0.98\linewidth]{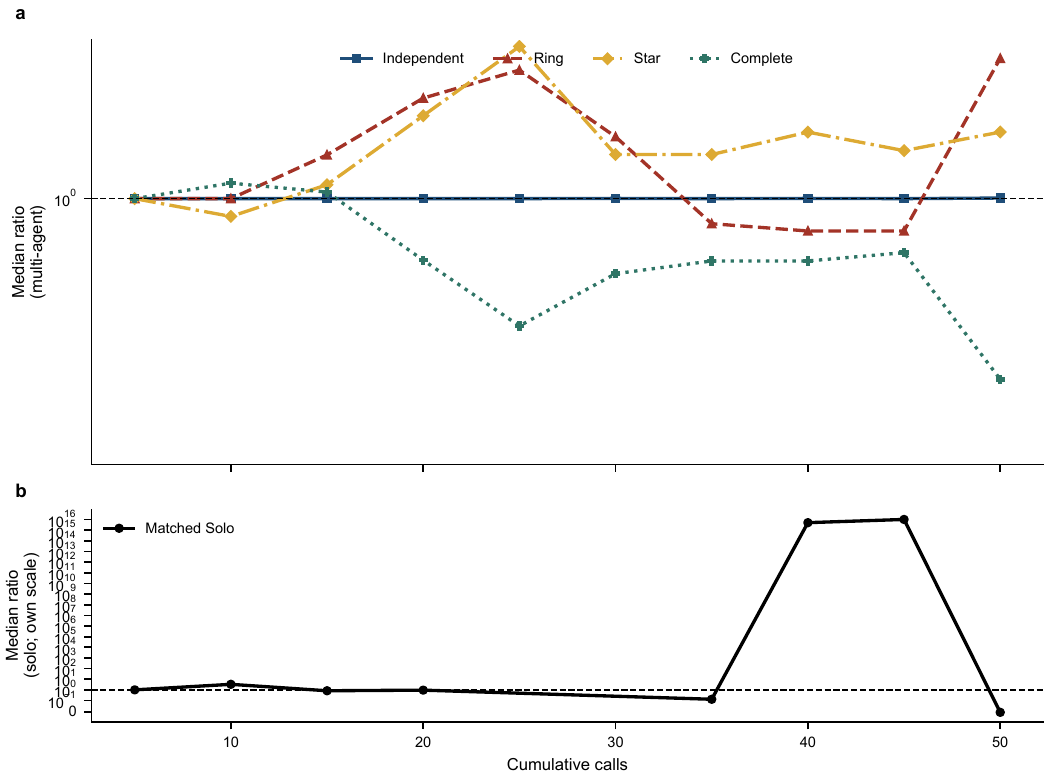}
\caption{Exploratory normalized symbolic-to-function dispersion ratio. Solo's extreme ratios result from positive function-dispersion denominators near numerical zero; the existing analysis excludes zero/nonfinite denominators as undefined but retains finite near-zero cases without clipping, so these magnitudes are numerically ill-conditioned, not evidence of a large effect.}
\label{fig:decoupling}
\end{figure}

\begin{figure}[htbp]
\centering
\includegraphics[width=0.98\linewidth]{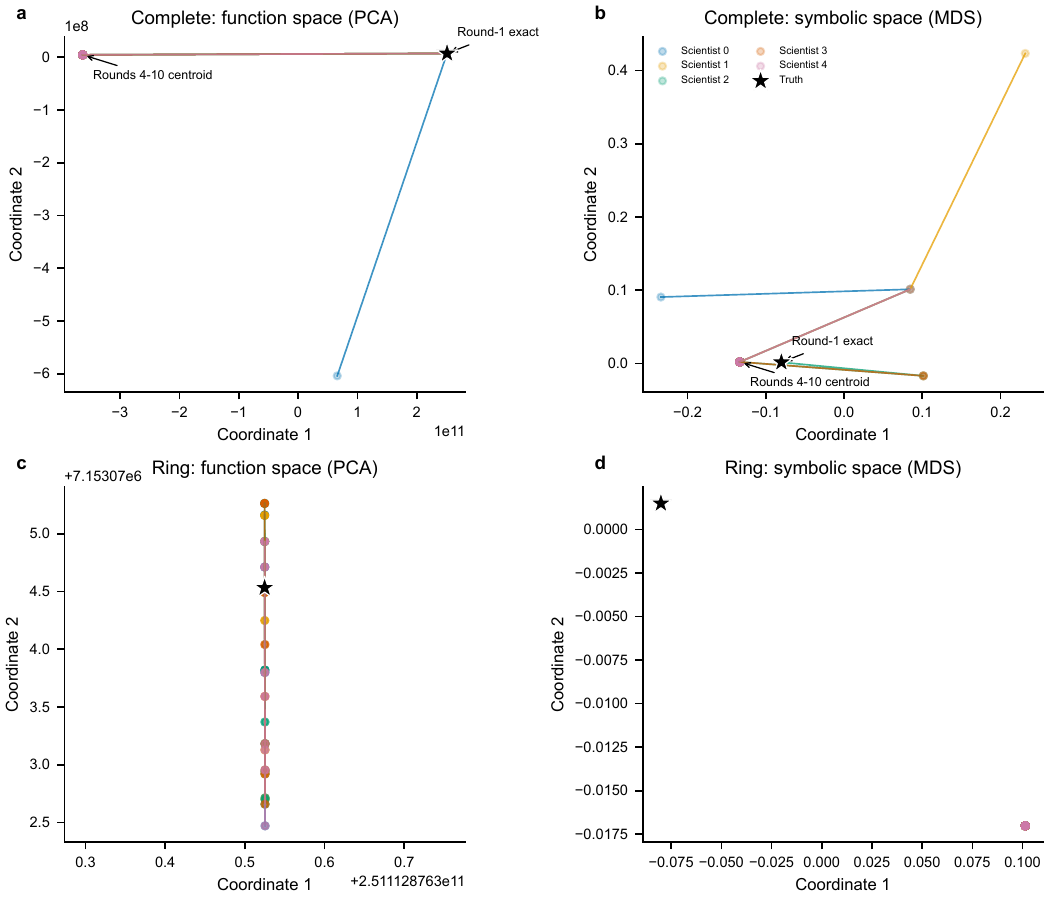}
\caption{Illustration for I.44.4\_4\_0 (Problem 34): Complete and Ring in function and symbolic spaces. PCA and MDS are fitted on this problem only; axes have no across-problem meaning. The reference is marked separately. This is a single-task illustration, not a statistical conclusion.}
\label{fig:problem34}
\end{figure}

\clearpage\FloatBarrier\setcounter{table}{0}\setcounter{figure}{0}
\section{Terra within-layer comparisons}\label{app:I}

The inference set is the common 41-problem Transform intersection. Each row below retains the reported mean interval and raw and Holm-adjusted signed-rank p values. Model settings differ from Qwen, and no model-by-policy interaction was tested. All medians are zero; the high frequency of ties is therefore part of the interpretation, not evidence that every difference vanishes.

\begingroup\tablefont
\setlength{\tabcolsep}{2.0pt}
\begin{longtable}{@{}>{\raggedright\arraybackslash}p{85.900pt}>{\raggedright\arraybackslash}p{106.442pt}>{\raggedright\arraybackslash}p{29.878pt}>{\raggedright\arraybackslash}p{37.348pt}>{\raggedright\arraybackslash}p{28.011pt}>{\raggedright\arraybackslash}p{28.011pt}>{\raggedright\arraybackslash}p{57.889pt}@{}}
\toprule
\textbf{Contrast} & \textbf{Mean delta Q [95\% CI]} & \textbf{Median} & \textbf{W/L/T} & \textbf{Raw p} & \textbf{Holm p} & \textbf{Family} \\
\midrule
\endfirsthead
\toprule
\textbf{Contrast} & \textbf{Mean delta Q [95\% CI]} & \textbf{Median} & \textbf{W/L/T} & \textbf{Raw p} & \textbf{Holm p} & \textbf{Family} \\
\midrule
\endhead
\midrule\multicolumn{7}{r}{\emph{Continued on next page}}\\\endfoot
\bottomrule\endlastfoot
Independent - Solo & +1.04 [0.06, 2.07] & 0 & 18/3/20 & \ensuremath{1.87\times10^{-3}} & \ensuremath{1.87\times10^{-3}} & Allocation, 1 test \\
Ring - Independent & -1.18 [-2.04, -0.48] & 0 & 2/19/20 & $5.43\!\times\!10^{-4}$ & \ensuremath{1.63\times10^{-3}} & Communication, 3 tests \\
Star - Independent & -0.57 [-1.56, +0.40] & 0 & 3/14/24 & 0.01 & 0.03 & Communication, 3 tests \\
Complete - Independent & -0.36 [-1.43, +0.73] & 0 & 4/13/24 & 0.08 & 0.08 & Communication, 3 tests \\
Ring - Star & -0.61 [-1.66, +0.30] & 0 & 6/13/22 & 0.06 & 0.13 & Policy, 3 tests \\
Ring - Complete & -0.82 [-1.93, +0.15] & 0 & 4/13/24 & 0.04 & 0.11 & Policy, 3 tests \\
\end{longtable}\endgroup

Star's signed-rank result passes Holm correction, while its mean interval includes zero. Both are retained because they address different features of the paired data. Appendix~\ref{app:D} supplies the full Terra delivery matrix and interaction family, and Appendix~\ref{app:J} records why Terra Independent on Synth was not run.

\begin{figure}[htbp]
\centering
\includegraphics[width=0.98\linewidth]{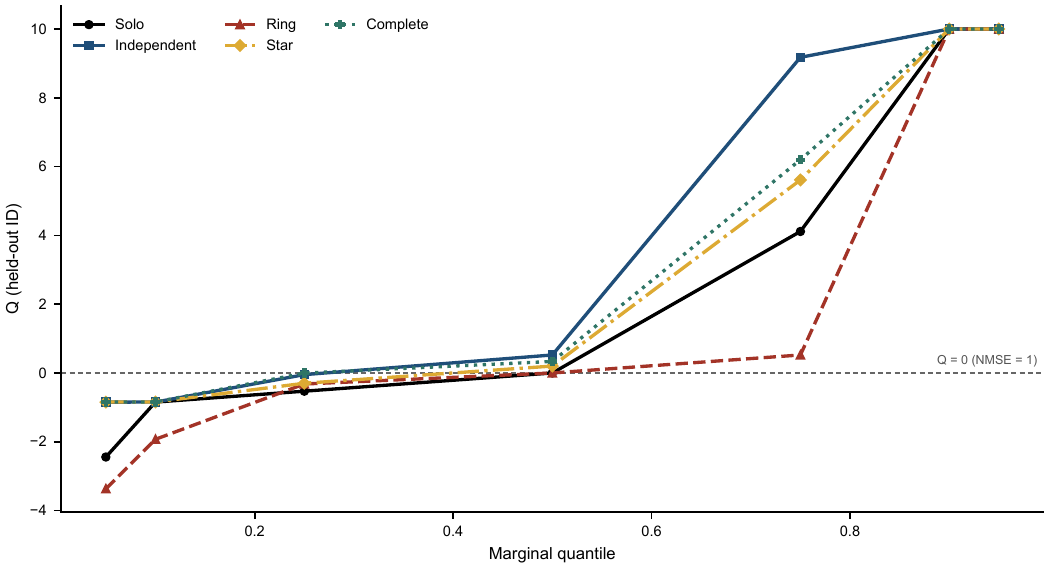}
\caption{Descriptive Terra held-out ID quantiles on the common 41-task Transform cohort. Q=0 corresponds approximately to NMSE=1. This restricted layer does not establish robustness on Synth.}
\label{fig:terra-quantiles}
\end{figure}

\clearpage\FloatBarrier\setcounter{table}{0}\setcounter{figure}{0}
\section{Analysis provenance and execution record}\label{app:J}

\subsection{Analysis status and endpoint history}

The three communication contrasts against Independent are exploratory. The earliest document explicitly defining this family was created on July 26, 2026 at 12:04:47, after the recorded \texttt{pilot\_complete/results.json} timestamp of 11:53:17 on the same date. Thus, the contrasts were explicitly specified after all five original conditions had produced results. We do not describe this family as confirmatory or as specified before observing the original results. The focal status of d2, d3, and d4 is also exploratory, despite the specified six-rule delivery family. No analysis in this manuscript is labeled \texttt{original\_primary}.

On August 28, 2026 at 00:28:13-05:00, before Analysis Protocol v2 Task 0 and reanalysis, the full-pilot primary endpoint was amended from OOD \(Q\) to pregenerated held-out ID \(Q\). All 50 selected Transform tasks structurally lack \texttt{ood\_test}. The original OOD endpoint was retained as a restricted endpoint on the 50 selected Synth tasks with that array. This amendment occurred before that reanalysis, not before the original experiments.

\subsection{Wilcoxon input convention and the completed sensitivity check}

The historical Transform analysis removed exact-zero paired differences before passing its input to SciPy. The later full-Qwen, allocation-sweep, and Terra analyses passed the full paired vector with \texttt{zero\_method=\textquotesingle{}wilcox\textquotesingle{}}. Although both conventions discard zero contributions to signed ranks, SciPy's automatic method selection can depend on whether zeros were present in the input. We retain the historical Transform values as the main reported values and show the full-vector sensitivity below. The historical SciPy version is unrecoverable.

\begingroup\tablefont
\setlength{\tabcolsep}{3.0pt}
\begin{longtable}{@{}>{\raggedright\arraybackslash}p{104.574pt}>{\raggedright\arraybackslash}p{67.226pt}>{\raggedright\arraybackslash}p{67.226pt}>{\raggedright\arraybackslash}p{67.226pt}>{\raggedright\arraybackslash}p{67.226pt}@{}}
\toprule
\textbf{Transform contrast} & \textbf{Historical raw $p$} & \textbf{Historical Holm $p$} & \textbf{Full-vector raw $p$} & \textbf{Full-vector Holm $p$} \\
\midrule
\endfirsthead
\toprule
\textbf{Transform contrast} & \textbf{Historical raw $p$} & \textbf{Historical Holm $p$} & \textbf{Full-vector raw $p$} & \textbf{Full-vector Holm $p$} \\
\midrule
\endhead
\midrule\multicolumn{5}{r}{\emph{Continued on next page}}\\\endfoot
\bottomrule\endlastfoot
Independent - matched Solo & $2.31\!\times\!10^{-4}$ & $2.31\!\times\!10^{-4}$ & $3.76\!\times\!10^{-4}$ & $3.76\!\times\!10^{-4}$ \\
Ring - Independent & $7.84\!\times\!10^{-5}$ & $2.35\!\times\!10^{-4}$ & $1.53\!\times\!10^{-4}$ & $4.60\!\times\!10^{-4}$ \\
Star - Independent & 0.29 & 0.29 & 0.28 & 0.28 \\
Complete - Independent & 0.11 & 0.21 & 0.11 & 0.21 \\
Ring - Star & $2.29\!\times\!10^{-6}$ & $4.57\!\times\!10^{-6}$ & $1.04\!\times\!10^{-5}$ & $2.09\!\times\!10^{-5}$ \\
Ring - Complete & $4.36\!\times\!10^{-7}$ & $1.31\!\times\!10^{-6}$ & $3.17\!\times\!10^{-6}$ & $9.50\!\times\!10^{-6}$ \\
Star - Complete & 0.91 & 0.91 & 0.91 & 0.91 \\
\end{longtable}\endgroup

The sensitivity check changes neither the effect directions nor the reported significance decisions. The one-comparison allocation family has a Holm value equal to its raw value; filling an earlier \texttt{null} metadata field did not change that test.

\subsection{Bootstrap interval correction}

An earlier draft combined the stated seed \texttt{20260904} with pooled Star and Complete intervals copied from the seed-\texttt{20260828} analysis. The present manuscript uses the authoritative corrected intervals: Star minus Independent, \([-0.08,\,+0.13]\); Complete minus Independent, \([-0.08,\,+0.34]\). Ring minus Independent retains \([-0.49,\,-0.18]\). Both Star and Complete intervals continue to span zero. The correction concerns interval provenance and does not convert an inconclusive comparison into evidence of equivalence.

\subsection{Execution changes and incomplete conditions}

\begingroup\tablefont
\setlength{\tabcolsep}{3.0pt}
\begin{longtable}{@{}>{\raggedright\arraybackslash}p{97.870pt}>{\raggedright\arraybackslash}p{293.610pt}@{}}
\toprule
\textbf{Item} & \textbf{Execution record and treatment in the analysis} \\
\midrule
\endfirsthead
\toprule
\textbf{Item} & \textbf{Execution record and treatment in the analysis} \\
\midrule
\endhead
\midrule\multicolumn{2}{r}{\emph{Continued on next page}}\\\endfoot
\bottomrule\endlastfoot
API constraints & On August 26, GPT-5-family probes rejected \nolinkurl{max_tokens} and temperature 0.70. The runner switched these requests to \nolinkurl{max_completion_tokens}; the Terra layer used temperature 1.00 and \nolinkurl{reasoning_effort=none}. Model, reasoning setting, and temperature are treated as a separate layer. \\
Matched Solo & The July Solo condition had 1 x 10=10 calls per problem, versus 5 x 10=50 in the five-agent conditions. \nolinkurl{pilot_solo_matched}, launched August 26 and completed August 27, used 1 x 50=50 and superseded the July Solo run for fixed-budget inference. Historical outputs remain preserved. \\
N=2 continuation & The first request for manifest index 130, BPG5, timed out after 69 completed problems and 3,450 persisted calls. The exact-parameter resume skipped those 69 completed problems, restarted index 130 from its first call, and completed all 100 problems. The timed-out request produced no usable model output. \\
Terra Synth & Terra Solo Synth completed 50 tasks, but the Independent Synth run stopped on its first request with HTTP 429 \nolinkurl{credit_balance_exhausted}: 0/50 completed problems, zero persisted successful calls, and USD 0.00. No result is imputed. \\
Post-hoc diversity analysis & A prespecified stopping rule on the pooled paired correlation was triggered on September 5 and the downstream analyses were not run. The stratified test was then executed as a recorded deviation, with the reason and the fixed assignment rule stated in Appendix G.3. No new model calls were made; the frozen archives and their aggregate SHA-256 were unchanged. \\
Removed self-reproduction statistic & The multi-agent lag-5 result was removed from inference on September 4. For length-10 sequences, the approved circular-shift null is invariant at lag 5 and produces a degenerate point-mass null; $p=1$ is not informative. Multi-agent lags 1–4 remain available, and Solo lags 1–10 are descriptive. \\
Symbolic similarity fallback & Exact tree edit distance could not be computed because \nolinkurl{zss} and \nolinkurl{apted} were unavailable. The documented fallback parses frozen canonical skeletons with SymPy, counts operator and variable labels as multisets, and calculates one minus multiset-intersection size divided by multiset-union size. These results are exploratory. \\
Original-run source guard & The frozen original-run manifest covers 1,341 files. The recorded checks before and after verification passed with zero mismatches. The complete tree SHA-256 is \nolinkurl{6e8874159b78ebadc5c48037017a9ce1a89f043c58d7d0a43ed0bcfe34fcf532}. \\
\end{longtable}\endgroup

The evidence for these records is identified in the data-completion JSON under \texttt{direct\_replacements.appendix\_j\_execution\_rows}, including the runner, configuration, execution-log, journal, and verification-report references. The completion audit made zero model API calls and did not modify the original-run tree.

\subsection{Accompanying numerical source register}

The accompanying source package contains the following authority exports in its \texttt{data/} directory. The delivery CSVs retain each row's exact valid-problem identifiers. Display rounding in this manuscript does not replace these full-precision records.

\begin{itemize}
\tightlist
\item
  Qwen simple effects: \nolinkurl{2026-09-05-qwen-delivery-simple-effects.csv}.
\item
  Qwen delivery interactions: \nolinkurl{2026-09-05-qwen-delivery-interactions.csv}.
\item
  Terra simple effects: \nolinkurl{2026-09-05-terra-delivery-simple-effects.csv}.
\item
  Terra delivery interactions: \nolinkurl{2026-09-05-terra-delivery-interactions.csv}.
\item
  Qwen task-family contrasts: \nolinkurl{2026-09-05-qwen-stratified-contrasts.csv}.
\item
  Full statistical completion: \nolinkurl{2026-09-04-part9-completion.json}.
\item
  Manuscript-data audit and provenance: \nolinkurl{2026-09-05-manuscript-v1-data-completion.json}.
\item
  Candidate-diversity archive export: \nolinkurl{2026-09-05-diversity-analysis.json}.
\item
  P5 stratified-deviation export: \nolinkurl{2026-09-05-p5-stratified.json}.
\item
  Pooled diversity-correlation figure and source-data sidecar: \nolinkurl{fig-paired-deltaD-deltaQ.pdf} and \nolinkurl{fig-paired-deltaD-deltaQ.json}.
\item
  P5 stratified figure and source-data sidecar: \nolinkurl{fig-p5-stratified.pdf} and \nolinkurl{fig-p5-stratified.json}.
\end{itemize}

Figure PDFs used by the manuscript are retained in \texttt{figures/}; the release snapshot preserves the available analysis sidecars. This register identifies supporting records, not additional experiments or a new inferential family.